\documentclass[trackchanges,twocolumn]{aastex701}

\usepackage{amsmath}

\begin{document}

\title{Fast X-Ray Transients from Dense Circumstellar Material Shock Breakout and Cooling Emission}

\author[orcid=0000-0002-9195-4904]{Jin-Ping Zhu}
\affiliation{The Hong Kong Institute for Astronomy and Astrophysics, The University of Hong Kong, Pokfulam Road, Hong Kong, China}
\affiliation{Department of Physics, The University of Hong Kong, Pokfulam Road, Hong Kong, China}
\email[show]{\url{jpzhu@hku.hk}}

\author[orcid=0000-0002-5596-5059]{Yi-Han Iris Yin}
\affiliation{The Hong Kong Institute for Astronomy and Astrophysics, The University of Hong Kong, Pokfulam Road, Hong Kong, China}
\affiliation{Department of Physics, The University of Hong Kong, Pokfulam Road, Hong Kong, China}
\email{\url{iris.yh.yin@connect.hku.hk}}

\author[orcid=0000-0002-1067-1911]{Yun-Wei Yu}
\affiliation{Institute of Astrophysics, Central China Normal University, Wuhan 430079, China}
\affiliation{Laboratory for Compact Object Astrophysics and Astronomical Technology, Central China Normal University, Wuhan 430079, China}
\affiliation{Education Research and Application Center, National Astronomical Data Center, Wuhan 430079, China}
\email{\url{yuyw@ccnu.edu.cn}}

\author{Zhuo Li}
\affiliation{Department of Astronomy, School of Physics, Peking University, Beijing 100871, China}
\affiliation{Kavli Institute for Astronomy and Astrophysics, Peking University, Beijing 100871, China}
\email{\url{zhuo.li@pku.edu.cn}}

\author[orcid=0000-0002-9725-2524]{Bing Zhang}
\affiliation{The Hong Kong Institute for Astronomy and Astrophysics, The University of Hong Kong, Pokfulam Road, Hong Kong, China}
\affiliation{Department of Physics, The University of Hong Kong, Pokfulam Road, Hong Kong, China}
\email[show]{\url{bzhang1@hku.hk}}

\begin{abstract}

In this {\em{Letter}}, we explore the soft X-ray shock breakout (SBO) and cooling emission of supernovae interacting with pre-explosion dense circumstellar material (CSM) shells as a subclass of extragalactic fast X-ray transients. For geometrically thin CSM shells, we find that the SBO luminosities span a range of a few $10^{43}-10^{45}\,{\rm{erg}}\,{\rm{s}}^{-1}$, with peak times typically $\lesssim10^4\,{\rm{s}}$, both of which are primarily determined by the light-travel effect. When the shell's outer radius is located at $\sim100-500\,R_\odot$, the SBO temperatures fall into the energy band covered by the current soft X-ray facilities. We derive analytic lightcurves for SBOs in dense CSM shells and explore how different physical parameters affect the observed lightcurves and the temperature evolution. Despite the multi-temperature nature of the emission caused by the light-travel effect leading to non-thermal-like spectra, we find that the time-resolved spectra within the limited soft X-ray band can still be well described by a blackbody model. This provides a natural explanation for the observed blackbody-like time-resolved spectra of soft X-ray SBO events. Our model can reproduce the observed X-ray lightcurves and temperature evolutions of the three known X-ray SBO events, EP260321a, XRF\,060218, and XRO\,080109, revealing a diversity in the properties of the dense CSM shells ejected by stripped-envelope progenitors before the explosion. Future discoveries of more soft X-ray SBO events by facilities such as Einstein Probe will provide opportunities for probing the episodic mass-loss history of massive stars in the final stages before core collapse.

\end{abstract}

\keywords{\uat{Shocks}{2086} --- \uat{Circumstellar shells}{242} --- \uat{Supernovae}{1668} --- \uat{Light curves}{918} --- \uat{Massive stars}{732} }

\section{Introduction} \label{sec:Introduction}

Powerful shocks propagating in optically thick environments are mediated by radiation, forming radiation-mediated shocks (RMSs), in which photons are trapped by the high optical depth of the upstream material and diffuse through the shock transition layer \citep{Colgate1974,Weaver1976,Katz2010}. RMSs can arise in different environments, typically including the stellar interior during core-collapse explosions \citep[e.g.,][]{Matzner1999,Nakar2010,Sapir2011} and dense circumstellar material (CSM) surrounding the exploding progenitor star, such as stellar winds \citep[e.g.,][]{Ofek2010,Chevalier2011,Balberg2011,Svirski2012,Svirski2014,Ioka2019,Khatami2024}, pre-ejected dense CSM shells \citep[e.g.,][]{Smith2007,Margalit2022}, and even AGN disks \citep[e.g.,][]{Zhu2021b,Zhu2021a,Perna2021,Grishin2021}. When an RMS reaches the stellar surface or the outer edge of the CSM, the trapped photons rapidly escape, producing a rapidly evolving and luminous shock breakout (SBO) flash. Such SBO events typically last from seconds to hours and can emerge across the UV to X-ray bands, with observed properties mainly determined by the progenitor radius, density structure, and shock properties \citep[see][for review]{Waxman2017,Levinson2020}. In particular, SBOs producing luminous soft X-ray flashes represent a promising class of extragalactic fast X-ray transients (EFXTs). However, due to the limited sensitivity and sky coverage of previous X-ray facilities, as well as the intrinsically rapid evolution of SBO emission, only two well-established soft X-ray SBO events, i.e., XRF\,060218 \citep{Campana2006} and XRO\,080109 \citep{Soderberg2008}, had been identified until recently. Both events were discovered in the early {\em Swift} era, with the prompt emission of the former being first triggered by the {\em Swift} Burst Alert Telescope (BAT), followed by the detection of its soft X-ray thermal SBO emission with {\em Swift} X-ray Telescope (XRT) observations, while the latter was serendipitously detected by {\em Swift}/XRT. Nevertheless, despite being regarded as soft X-ray SBO events, the physical nature of these events, including the shock velocity and breakout environment, remains under active debate \citep[e.g.,][]{Waxman2007,Couch2011,Svirski2014b,Nakar2015,Irwin2025}.

Historically, many EFXTs, characterized by short-lived X-ray emission lasting from seconds to hours, were serendipitously identified in archival observations from {\em Chandra} and {\em XMM-Newton} \citep{Jonker2013,Glennie2015,Bauer2017,Xue2019,Lin2022,QuirolaVasquez2022}, yet the physical origins of most of these events remain poorly understood, mainly due to the lack of follow-up observations. With the wide-field and high-sensitivity soft X-ray capabilities, the Einstein Probe \citep[EP;][]{Yuan2022} has enabled the detection of over 100 {EFXTs} since the mission launched, many of which have been found to be associated with explosive phenomena from massive stars \citep[e.g.,][]{LiuY2025,Sun2025,Jiang2025,Li2025,Cotter2026,Srinivasaragavan2026}. These so-called Type II EFXTs that are associated with massive star explosions can be theoretically classified into several subclasses according to their physical origins (see B. Zhang et al. 2026, in prep, for a review), successful strong dirty jets \citep{Dai2026b,Dai2026},  successful weak jets \citep{Sun2025,Hamidani2025,Li2025,Schneider2026}, gamma-ray bursts (GRBs) viewed off-axis \citep{Zheng2025,Zhu2025,Gianfagna2025,Zheng2026}, choked jets and relativistic SBOs, and non-relativistic SBOs. Most recently, \citet{Yuan2026} reported EP260321a, the first identified soft X-ray SBO event classified as a non-relativistic SBO EFXT, discovered in the EP era, which showed a peak luminosity of $(1.0\pm0.3)\times10^{45}\,{\rm erg}\,{\rm s}^{-1}$ and a peak time of $(690\pm60)\,{\rm s}$ after the trigger. The time-resolved spectra of EP260321a can be well fitted by blackbody spectra, with an inferred peak temperature of around $\sim0.1-0.13\,{\rm keV}$. \cite{Yuan2026} suggested that this event may originate from a supernova (SN) SBO from a thin dense shell, which can explain both the observed X-ray lightcurve and temperature evolution \citep[see also][who proposed a potential wind SBO origin but found it unable to explain the observed temperature evolution]{Suzuki2026}. 

{SBO and subsequent cooling from confined CSM shells have been investigated in several theoretical studies. Diffusion and cooling of shock-deposited energy in optically thick massive CSM shells have been proposed as a potential origin for powering peculiar transients, such as hydrogen-rich superluminous SNe and fast blue optical transients \citep[e.g.,][]{Smith2007,Rest2018}. Analytic models of shock cooling from extended material and dense CSM shells were developed by \cite{Piro2021,Margalit2022}, with the latter explicitly treating radiative diffusion and the transition from planar to spherical expansion. \cite{Khatami2024} mapped the interaction lightcurve regimes of confined CSM configurations. For CSM shells with optical depth $\tau_{\rm CSM}\lesssim c/v_{\rm sh}$ (where $v_{\rm sh}$ is the shock velocity), \cite{Margalit2022b} investigated collisionless-shock bremsstrahlung emission, including Comptonization, photoelectric absorption, and post-interaction cooling. Building on previous studies, in this {\em Letter}, we combine a non-LTE
color-temperature prescription, light-travel effect, and a
piecewise cooling lightcurve within a single fitting framework for
jointly modelling the soft X-ray luminosity and temperature evolution.}
We systematically investigate the SBO properties and subsequent
cooling emission from SNe interacting with dense CSM shells, assess this
scenario as a possible emission mechanism for a subclass of EFXTs and apply the framework to interpret the three currently known soft X-ray SBO events.

\section{Shock Breakout Properties of Supernova in Dense CSM} \label{sec:SBO}

\subsection{Radiation Mechanism}

We consider a spherically symmetric dense CSM shell of mass $M_{\rm CSM}$ ejected by the progenitor before the SN explosion, characterized by an inner radius $r_{\rm in}$ and an outer radius $r_{\rm out}$. Both radii are assumed to be much larger than the progenitor radius $r_\star$. The dense CSM shell is taken to have a uniform density, $\rho_{\rm CSM} \approx 3M_{\rm CSM}/4\pi(r_{\rm out}^3-r_{\rm in}^3)$. This simplification does not significantly affect the resulting SBO and cooling emission, as a truncated wind density profile ($\rho_{\rm CSM}\propto r^{-2}$) gives similar results in the thin-shell limit. In this section, we explore the basic SBO properties in a thin-shell limit, where the CSM density is then approximated as $\rho_{\rm CSM}\sim M_{\rm CSM}/(4\pi r_{\rm out}^3\chi)$ with $\chi \equiv (r_{\rm out}-r_{\rm in})/r_{\rm out}\ll1$ being the dimensionless shell thickness parameter. A spherical SN ejecta subsequently collides with the CSM shell, leading to a powerful radiation-dominated shock propagating at a velocity $v_{\rm sh}$.

The SBO occurs when the effective diffusion speed matches the shock velocity (i.e., $v_{\rm sh}\sim c/\tau_{\rm SBO}$), or equivalently when the photon diffusion timescale ($t_{\rm diff,SBO}\approx\tau_{\rm SBO}\Delta r_{\rm SBO}/c$) becomes comparable to the shock dynamical timescale ($t_{\rm dyn,SBO}\approx\Delta r_{\rm SBO}/v_{\rm sh}$), where $\tau_{\rm SBO}\approx\kappa\rho_{\rm CSM}\Delta r_{\rm SBO}$ denotes the optical depth of the unshocked CSM ahead of the diffusion front at breakout, with $\kappa$ being the opacity and the thickness thus given by $\Delta r_{\rm SBO}\approx c/\kappa\rho_{\rm CSM}v_{\rm sh}$. {As observed soft X-ray SBOs are associated with stripped-envelope SNe, we adopt $\kappa \approx 0.2\,{\rm cm^2\,g^{-1}}$ for fully ionized He/C/O-rich material.} The radiation energy of the SBO originates from the thermal energy stored in the radiation-dominated layer near the shock front, which is suddenly released and then deposited into the upstream material. The corresponding SBO energy, which originates from the kinetic energy dissipated by the shock in the material outside the breakout layer, can be estimated as
\begin{equation}
\begin{split}
    E_{\rm SBO}&\approx \frac{1}{2}m_{\rm SBO} v_{\rm sh}^2\approx \frac{2\pi cr_{\rm out}^2v_{\rm sh}}{\kappa}\\
    &= 3\times10^{47}\,\kappa_{-0.7}^{-1}r_{\rm out,13}^2\left(\frac{v_{\rm sh}}{0.1\,c}\right)\,{\rm erg},
\end{split}
\end{equation}
where the total mass of the upstream CSM ahead of the shock is $m_{\rm SBO}\approx 4\pi \rho_{\rm CSM}r_{\rm out}^2\Delta r_{\rm SBO}=3.2\times10^{-5}\,\kappa_{-0.7}^{-1}r_{\rm out,13}^2(v_{\rm sh}/0.1\,c)^{-1}\,M_\odot$. Hereafter, the conventional notation $Q_x=Q/10^x$ is adopted in cgs units. 

The observed peak of an SBO signal can be determined by both photon diffusion and light-travel time. If the peak is dominated by photon diffusion, the peak timescale is equivalent to both the diffusion timescale and the dynamical timescale, i.e., $t_{\rm SBO}\approx t_{\rm diff,SBO}\approx t_{\rm dyn,SBO}$. In this case, the observed SBO luminosity is
\begin{equation}
\begin{split}
\label{equ:Luminosity_Diffusion}
    &L_{\rm SBO}\sim\frac{E_{\rm SBO}}{t_{\rm diff,SBO}}\approx2\pi r_{\rm out}^2 \rho_{\rm CSM}v_{\rm sh}^3\sim\frac{M_{\rm CSM}v_{\rm sh}^3}{2r_{\rm out}\chi} \\
    &= 2.7\times10^{45}\,r_{\rm out,13}^{-1}\chi_{-1}^{-1}\left(\frac{M_{\rm CSM}}{10^{-4}\,M_\odot}\right)\left(\frac{v_{\rm sh}}{0.1\,c}\right)^3\,{\rm erg}\,{\rm s}^{-1},
\end{split}
\end{equation}
which corresponds to the instantaneous shock heating luminosity at the SBO. If the peak is dominated by the light-travel effect, the peak timescale is $t_{\rm SBO}\approx t_{\rm lt} \approx r_{\rm out}/c$, and the observed SBO luminosity is then given by 
\begin{equation}
\begin{split}
\label{equ:Luminosity_LightTravel}
    &L_{\rm SBO}\sim\frac{E_{\rm SBO}}{t_{\rm lt}}\approx \frac{2\pi c^2 r_{\rm out}v_{\rm sh}}{\kappa}\\
    &=8.5\times10^{44}\,\kappa_{\rm -0.7}^{-1}r_{\rm out,13}\left(\frac{v_{\rm sh}}{0.1\,c}\right)\,{\rm erg}\,{\rm s}^{-1}.
\end{split}
\end{equation}
The transition of the SBO peak between the diffusion-dominated and light-travel-time-dominated regimes is determined by equating Equations (\ref{equ:Luminosity_Diffusion}) and (\ref{equ:Luminosity_LightTravel}). 

If local thermal equilibrium is achieved at the shock diffusion front, the emergent spectrum approaches a blackbody, and the observed color temperature coincides with the equilibrium blackbody temperature, i.e.,
\begin{equation}
\label{equ:BBTemperature}
    T_{\rm BB} \sim({\epsilon}/{a})^{1/4}
\end{equation}
where $\epsilon\sim E_{\rm SBO}/4\pi r_{\rm out}^2\Delta r_{\rm SBO}$ is the photon energy density for all materials ahead of the diffusion front at the SBO, and $a$ is the radiation constant. However, when $v_{\rm sh}\gtrsim0.05\,c$, free-free emission cannot generate sufficient photons to maintain local thermal equilibrium \citep{Nakar2010,Katz2010}, while photons and electrons might reach Compton equilibrium with an observed color temperature $T_{\rm col}$ higher than the equilibrium blackbody temperature $T_{\rm BB}$. One can introduce a thermal-coupling parameter,
\begin{equation}
\begin{split}
\label{equ:eta}
    &\eta \approx\frac{n_{\rm BB}}{t_{\rm diff}\dot{n}_{\rm ph,ff}(T_{\rm BB})}\\
    &\approx\frac{7\times10^5\,{\rm s}}{t_{\rm diff}}\left(\frac{\rho_{\rm sh}}{10^{-10}\,{\rm g}\,{\rm cm}^{-3}}\right)^{-2}\left(\frac{k_{\rm B}T_{\rm BB}}{100\,{\rm eV}}\right)^{7/2},
\end{split}
\end{equation}
where $k_{\rm B}$ is the Boltzmann constant, $n_{\rm BB}$ is the blackbody photon number density, $\dot{n}_{\rm ph,ff}$ is the free-free photon production rate, $t_{\rm diff}$ is the diffusion timescale for the diffusion front (i.e., $t_{\rm diff,SBO}$ at the SBO), and $\rho_{\rm sh}\simeq7\rho_{\rm CSM}$ is the post-shock density,  assuming a shock compression factor of $7$. The observed color temperature may be expressed as \citep[e.g.][]{Nakar2010}
\begin{equation}
\label{equ:ColorTemperature}
    T_{\rm col} = \begin{cases} T_{\rm BB}\eta^2(T_{\rm BB})/\xi^2(T_{\rm col}), & {\rm if}\ \eta>1,\\
        T_{\rm BB}, & {\rm if}\ \eta\leq1,\\
    \end{cases}
\end{equation}
where $\xi(T)\approx\max[1,(1/2)\ln y_{\rm max}(1.6+\ln y_{\rm max})]$ is the ratio of photons at energy $3k_{\rm B}T$ created by Comptonization relative to free-free
processes, and $y_{\rm max}=k_{\rm B}T/h\nu_{\rm min}\approx3(\rho_{\rm sh}/10^{-9}\,{{\rm g}\,{\rm cm}^{-3}})^{-1/2}(k_{\rm B}T/100\,{\rm eV})^{9/4}$ represents the ratio of the thermal energy scale $T$ to the lowest energy photon that can be Comptonized. 

As an example, for a shock velocity of $v_{\rm sh}\sim0.1\,c$ propagating through a $10^{-4}M_\odot$ CSM shell with an outer radius of $r_{\rm out}=10^{13}\,{\rm cm}$ and a dimensionless shell thickness of $\chi=0.1$, the equilibrium blackbody temperature at the SBO is derived as $T_{\rm BB,SBO}\approx 0.05\,r_{\rm out,13}^{-3/4}\chi_{\rm -1}^{-1/4}(M_{\rm CSM}/10^{-4}\,M_\odot)^{1/4}(v_{\rm sh}/0.1\,c)^{1/2}\,{\rm keV}$, while the thermal-coupling parameter is $\eta\approx4\,\kappa_{-0.7}r_{\rm out,13}^{3/8}\chi_{-1}^{1/8}(M_{\rm CSM}/10^{-4}\,M_\odot)^{-1/8}(v_{\rm sh}/0.1\,c)^{15/4}$, indicating that the radiation is out of local thermal equilibrium. As a result, the observed color temperature is higher than the equilibrium blackbody temperature, with $T_{\rm col,SBO}\approx0.127\,{\rm keV}$, which falls in the soft X-ray band of the EP, and the corresponding value of $\xi(T_{\rm col,SBO})\approx2.5$.

\subsection{Shock Dynamics and Shock Breakout Emission Properties}

While the SBO observables depend on the CSM shell structure and the properties of the shock, the shock itself is not an independent input but is determined by the ejecta-CSM interaction, which couples the properties of both the SN ejecta and the CSM. We then describe how the shock properties are set by the ejecta-CSM interaction, taking into account both the CSM structure and the ejecta density profile.

The internal energy of the shocked CSM region is generated by the dissipation of the kinetic energy of the upstream flow through the forward shock, which can be estimated as $E_{\rm int,CSM}\approx M_{\rm CSM}v_{\rm sh}^2/2$. This energy is actually supplied by the kinetic energy of the high-velocity tail of the ejecta, characterized by a total mass $M_{\rm ej}$ and kinetic energy $E_{\rm kin}$\footnote{{We note that $E_{\rm kin}$ refers to the ejecta kinetic energy during CSM interaction, which may differ from SN-inferred values due to potential additional acceleration of central-engine injection \citep[e.g., from a magnetar; ][]{Zhu2026} after SBO.}}, with the relevant contribution dominated by ejecta components satisfying $v_{\rm ej}\gtrsim v_{\rm sh}$, such that $E_{\rm int,CSM}\sim E_{\rm kin}(>v_{\rm sh})$. Here, we assume that the SN ejecta has a broken-power-law density profile in velocity space $v$, with a flat profile in the inner region and a steep one in the outer region, which can be expressed as follows \citep{Chevalier1989,Matzner1999}
\begin{equation}
\label{equ:DensityProfile}
    \rho_{\rm ej}(v,t) = \left\{\begin{aligned}
 \zeta_\rho\frac{M_{\rm ej}}{v_{\rm tr}^3t^3}\left(\frac{v}{v_{\rm tr}} \right)^{-n},~v\geq v_{\rm tr}, \\
 \zeta_\rho\frac{M_{\rm ej}}{v_{\rm tr}^3t^3}\left(\frac{v}{v_{\rm tr}} \right)^{-\delta},~v< v_{\rm tr},
\end{aligned}\right.
\end{equation}
where $v_{\rm tr}\approx\zeta_\nu\sqrt{E_{\rm kin}/M_{\rm ej}}$ is the transition velocity, $\zeta_\rho=(n-3)(3-\delta)/4\pi(n-\delta)$, and $\zeta_\nu=\sqrt{2(5-\delta)(n-5)/(n-3)(3-\delta)}$. Then, the kinetic energy above the velocity $v$ is given by
\begin{equation}
\begin{split}
\label{equ:KineticProfile}
    E&_{\rm kin}(>v) = 2\pi\zeta_\rho M_{\rm ej}v_{\rm tr}^2\times\\
    &\begin{cases}
\dfrac{1}{n-5}\left(\dfrac{v}{v_{\rm tr}}\right)^{5-n}, & v \ge v_{\rm tr}, \\
\dfrac{1}{n-5}
+ \dfrac{1}{5-\delta}
\left[1-\left(\dfrac{v}{v_{\rm tr}}\right)^{5-\delta}\right], & v < v_{\rm tr}.
\end{cases}
\end{split}
\end{equation}
For the interaction only with the steep outer ejecta profile, i.e., $v_{\rm sh} \geq v_{\rm tr}\sim0.03\,E_{\rm kin,51}^{1/2}(M_{\rm ej}/2\,M_\odot)^{-1/2}c$, the shock velocity is given by
\begin{equation}
\begin{split}
\label{equ:ShockVelocity1}
    &v_{\rm sh}\sim v_{\rm tr}\left(\frac{4\pi\zeta_\rho}{5}\frac{M_{\rm ej}}{M_{\rm CSM}}\right)^{1/7}, \\
    &=0.1\,E_{\rm kin,51}^{1/2} \left(\frac{M_{\rm ej}}{2\,M_\odot}\right)^{-5/14}\left(\frac{M_{\rm CSM}}{10^{-4}\,M_\odot}\right)^{-1/7}c,
\end{split}
\end{equation}
where the fiducial values of $\delta=1$ and $n=10$ are adopted. If the CSM can interact with the flat inner ejecta core (i.e., $v_{\rm sh} < v_{\rm tr}$), it instead takes the form
\begin{equation}
\label{equ:ShockVelocity2}
    v_{\rm sh}\sim v_{\rm tr}\left(\frac{\sqrt{M_{\rm CSM}^2+36\pi^2\zeta_\rho^2 M_{\rm ej}^2/5}-M_{\rm CSM}}{2\pi\zeta_\rho M_{\rm ej}}\right)^{1/2}.
\end{equation}
We note that a radiation-mediated shock can form when $v_{\rm sh}\gtrsim c/\tau_{\rm CSM}$, where $\tau_{\rm CSM}\sim \kappa \rho_{\rm CSM}\Delta r_{\rm CSM}\sim\kappa M_{\rm CSM}/4\pi r_{\rm out}^2$ is the diffusion depth of the CSM shell, with $\Delta r_{\rm CSM}$ being the shell thickness. The condition for sustaining a radiation-mediated shock is
\begin{equation}
\begin{split}
    &M_{\rm CSM}\gtrsim\left[\frac{4\pi r_{\rm out}^2c}{\kappa v_{\rm tr}}\left(\frac{5}{4\pi\zeta_\rho M_{\rm ej}}\right)^{1/7}\right]^{7/6} \\
    &\gtrsim 3\times10^{-5}\,\kappa_{-0.7}^{-7/6}r_{\rm out,13}^{7/3}E_{\rm kin,51}^{-7/12}\left(\frac{M_{\rm ej}}{2\,M_\odot}\right)^{5/12}\,M_\odot,
\end{split}
\end{equation}
where the second expression assumes that SBO occurs in the outer ejecta envelope.

\begin{figure}[t!]
\centering
\includegraphics[width=1\linewidth, trim = 36 92 55 60, clip]{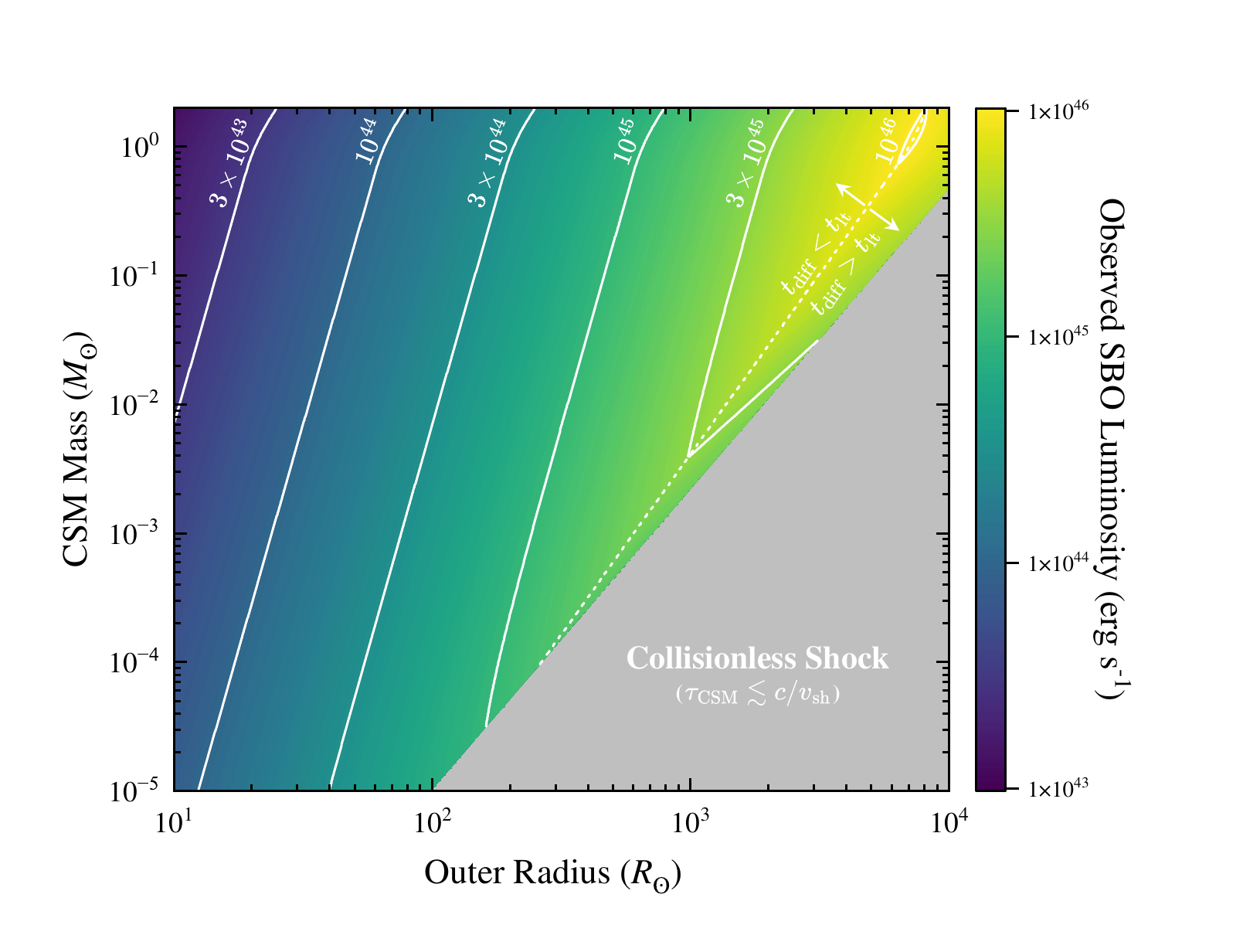}
\includegraphics[width=1\linewidth, trim = 36 92 55 60, clip]{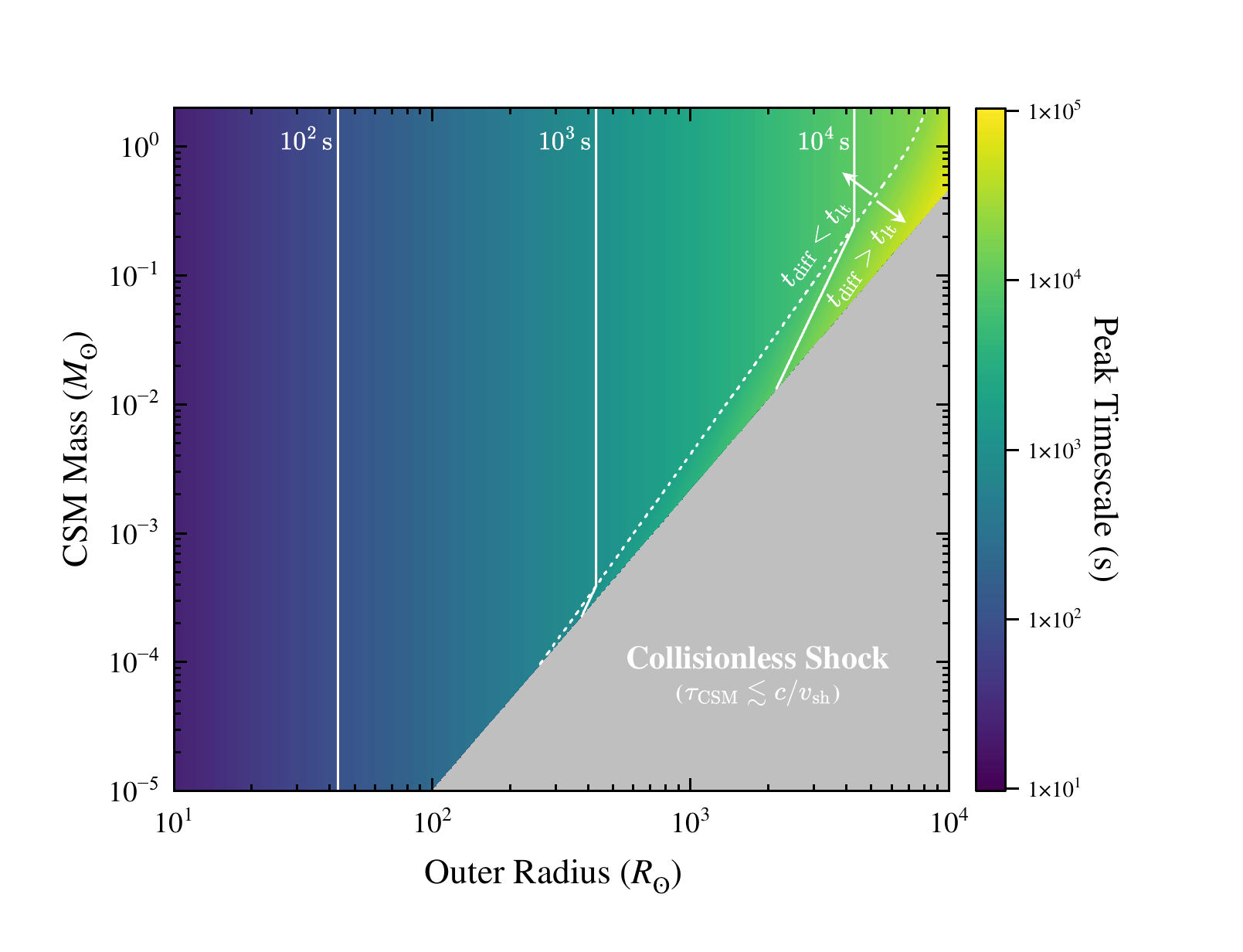}
\includegraphics[width=1\linewidth, trim = 36 34 55 60, clip]{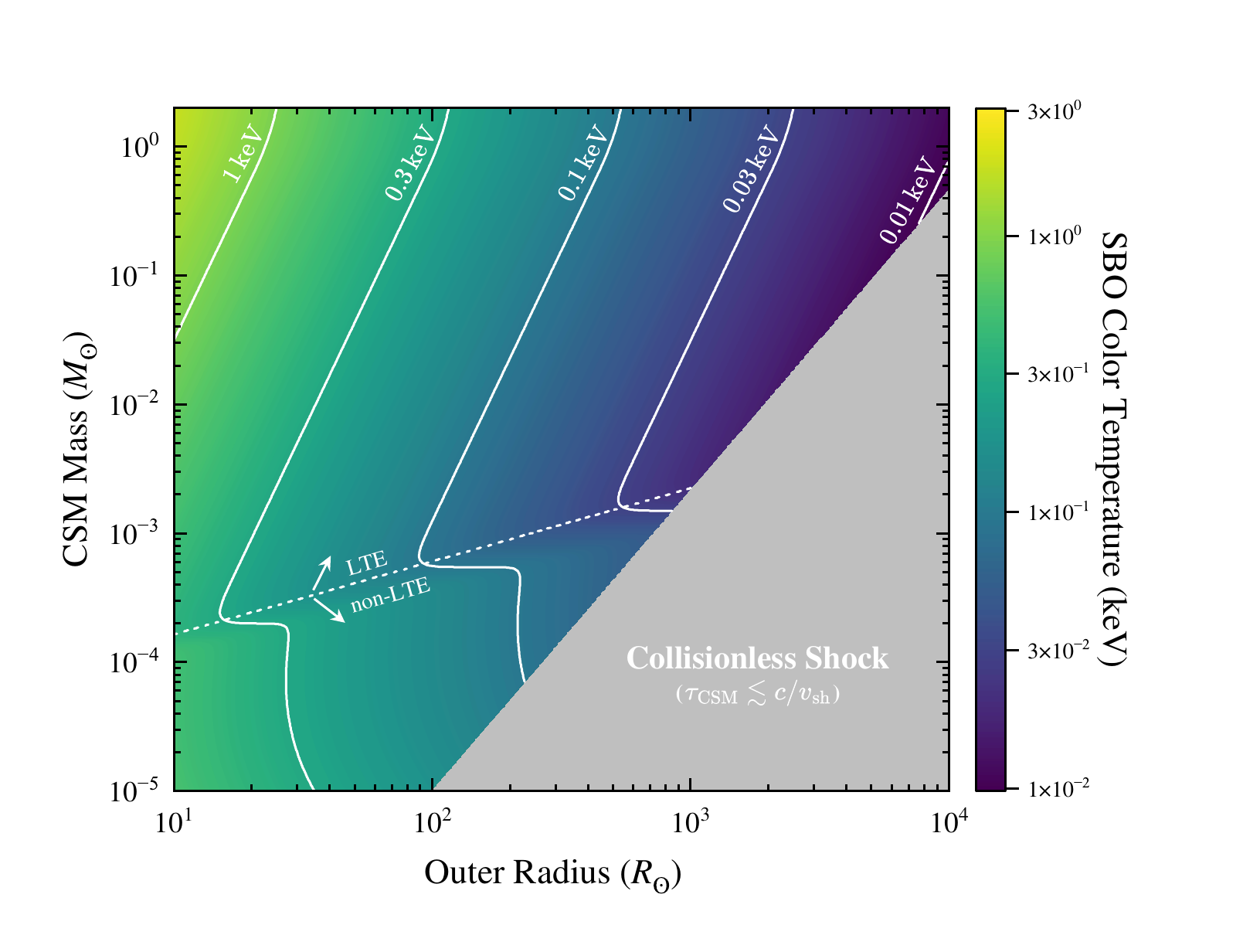}
\caption{Observed SBO properties, including the observed SBO Luminosity $L_{\rm SBO}$ (top), peak timescale $t_{\rm SBO}$ (middle), and color temperature $T_{\rm col,SBO}$ (bottom), in the $r_{\rm out}-M_{\rm CSM}$ plane. The colored region indicates the parameter space where a radiation-mediated shock can be formed and produce an observable SBO, while the gray region corresponds to cases where SBO cannot occur. In the top and middle panels, the dashed line separates the light-travel dominated regime ($t_{\rm diff}<t_{\rm lt}$; left) from the photon-diffusion dominated regime ($t_{\rm diff}>t_{\rm lt}$; right) for the observed luminosity and duration. In the bottom panel, the dashed line separates the LTE SBO regime ($\eta\leq1$; above) from the non-LTE regime ($\eta>1$; below). Here, $\chi=0.1$, $M_{\rm ej}=2\,M_\odot$, and $E_{\rm kin}=10^{51}\,{\rm erg}$ are fixed. }
\label{fig:SBOProperites}
\end{figure}

Assuming a typical SN ejecta with $M_{\rm ej}=2\,M_\odot$ and $E_{\rm kin}=10^{51}\,{\rm erg}$ interacting with a dense CSM characterized by $r_{\rm out}\sim10^1-10^4\,R_\odot$ and $M_{\rm CSM}\sim10^{-5}-2\,M_\odot$, we explore the parameter space of the observed SBO properties, including $L_{\rm SBO}$, $t_{\rm SBO}$, and $T_{\rm col,SBO}$ in Figure \ref{fig:SBOProperites}. Here, the dimensionless shell thickness is fixed at $\chi=0.1$ as a fiducial value, as the SBO observables are relatively insensitive to $\chi$ (see Section \ref{sec:ParameterDependence} for details on parameter dependence by $\chi$). The observed SBO properties show a clear dependence on the CSM properties, where more massive and more extended CSM shells lead to higher SBO luminosities and longer durations, but result in lower observed color temperatures. For $\chi=0.1$, $L_{\rm SBO}$ and $t_{\rm SBO}$ are mainly determined by the light-travel effect, with diffusion becoming relevant only for massive and extended CSM. As $\chi$ increases, the lower CSM density is expected to delay the SBO and extend the diffusion timescale, thereby reducing the parameter space dominated by the light-travel effect. For CSM shells with $M_{\rm CSM}\lesssim10^{-4}-10^{-3}\,M_\odot$, the shock downstream of the SBO can deviate from LTE. We note that the flat temperature behavior is insensitive to $M_{\rm CSM}$ near the transition between LTE and non-LTE, where $\eta>1$ but $\xi\sim1$, indicating that the free-free process can no longer produce photons at a sufficient rate, while Comptonization cannot compensate for the deficit. For SBO events within the energy range of interest for EP ($T_{\rm col,SBO}\sim0.1-1\,{\rm keV}$), the outer radius of the CSM is $r_{\rm out}\lesssim100-500\,R_\odot$. In this regime, the observed peak luminosity is generally regulated by the light-travel effect rather than photon diffusion.

\section{Shock Breakout and Cooling Lightcurve} \label{sec:Model}

\subsection{Source Lightcurve and Temperature Evolution}

\begin{figure}[t!]
\centering
\includegraphics[width=1\linewidth, trim = 0 0 0 0, clip]{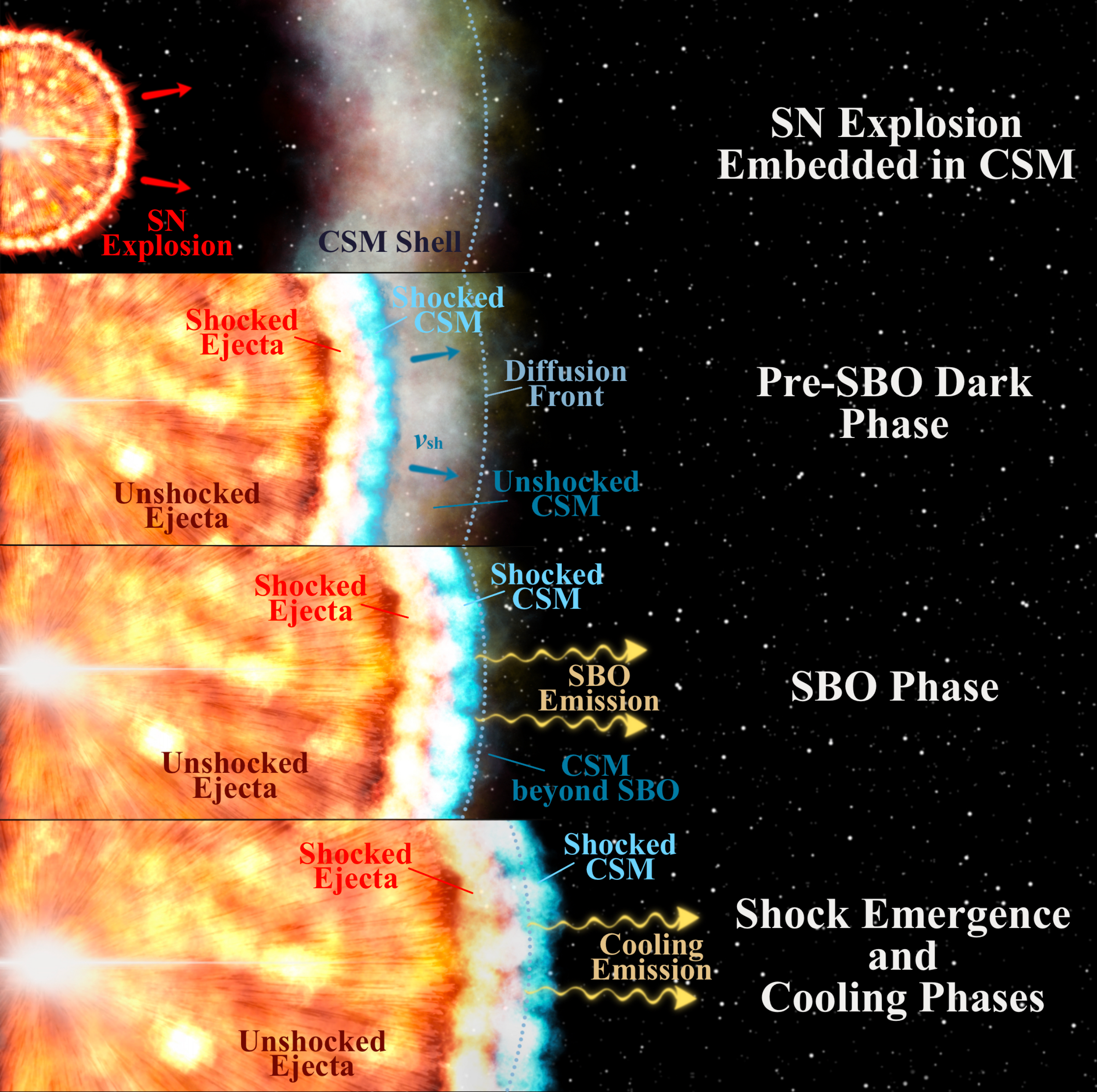}
\caption{Schematic illustration of a SN occurring within pre-SN CSM. }
\label{fig:Cartoon}
\end{figure}

In the context of soft X-ray SBO from a dense CSM shell as illustrated in Figure \ref{fig:Cartoon}, we divide the lightcurve evolution into three different phases: (0) the {\em Dark phase}, during which photons remain trapped behind the shock, preventing radiation from escaping to the observer; (1) the {\em SBO phase}, where the trapped shock energy is rapidly released when $\tau_{\rm SBO}\sim c/v_{\rm sh}$, leading to a sharp luminosity rise; (2) the {\em Shock Emergence Phase}, where the shock rapidly traverses the outer edge of the unshocked CSM; (3) the {\em Cooling Phase}, where the remaining shock-heated energy inside the shocked CSM is gradually released as the diffusion front propagates inward. Here, we only consider the emission from the shocked CSM, which produces the soft X-ray SBO and cooling emission of interest, and ignore the lightcurve contribution from the cooling of the shocked ejecta. The shocked CSM is assumed to be confined within a geometrically thin shell. We then first model the intrinsic source bolometric lightcurve and temperature evolution, before considering the light-travel effect. 

(1) {\em SBO Phase.} At the moment of the SBO, which we define as the zero point of time, i.e., $t=t_0$, the location of the SBO radius is 
\begin{equation}
    r_{\rm SBO} = r_{\rm out}-\Delta r_{\rm SBO}\approx r_{\rm out} - \frac{c}{\kappa\rho_{\rm CSM}v_{\rm sh}},
\end{equation}
where $r_{\rm SBO}$ should be larger than $r_{\rm in}$, implying the condition $v_{\rm sh}>c/\kappa\rho_{\rm CSM}(r_{\rm out}-r_{\rm in})$. The timescale $t_0$ is comparable to both $t_{\rm diff,SBO}$ and $t_{\rm dyn,SBO}$ introduced in Section \ref{sec:SBO}, which can be expressed as $t_{0}\sim\kappa\rho_{\rm CSM}(r_{\rm out}-r_{\rm SBO})^2/c\sim(r_{\rm out}-r_{\rm SBO})/v_{\rm sh}$. One can derive that the source luminosity at the SBO is
\begin{equation}
    L_{\rm sou}(t=t_0) \sim L_{\rm SBO}\approx\frac{M_{\rm CSM}(r_{\rm out}^3-r_{\rm SBO}^3)v_{\rm sh}^3}{2(r_{\rm out}^3-r_{\rm in}^3)(r_{\rm out}-r_{\rm SBO})},
\end{equation}
where $L_{\rm SBO}\sim E_{\rm SBO}/t_0$ is the SBO luminosity determined by the photon diffusion (i.e., Equation \ref{equ:Luminosity_Diffusion}), $E_{\rm SBO}\sim m_{\rm SBO}v_{\rm sh}^2/2$ is the SBO energy,  with $m_{\rm SBO}\approx 4\pi\rho_{\rm CSM}(r_{\rm out}^3-r_{\rm SBO}^3)/3$. The above equation can reduce to Equation (\ref{equ:Luminosity_Diffusion}) in the thin-shell limit for the CSM. For the rising phase at $t<t_0$, we adopt an empirical form for the source luminosity,
\begin{equation}
    L_{\rm sou}(t\leq t_0)\sim L_{\rm SBO}e^{1-t_0/t}
\end{equation}
which describes the diffusion of radiation leaking from the center of a static slab \citep[e.g.,][]{Nakar2010}.

(2) {\em Shock Emergence Phase.} During the shock crossing of the CSM edge, we define the shock position as $r_{\rm sh}\approx r_{\rm SBO}+v_{\rm sh}(t-t_0)\leq r_{\rm out}$, where $t_0<t\leq2t_0$. We locate the diffusion front by requiring the optical depth ahead of it to satisfy
\begin{equation}
\begin{split}
    \tau_{\rm diff}&\sim\tau_{\rm SC,diff}+\tau_{\rm UC,diff}\\
    &\sim\frac{\kappa m_{\rm SC,diff}}{4\pi r_{\rm sh}^2}+\kappa\rho_{\rm CSM}(r_{\rm out}-r_{\rm sh})\sim\frac{c}{v_{\rm sh}},
\end{split}
\end{equation}
where $\tau_{\rm SC,diff}$ and $\tau_{\rm UC,diff}$ are the optical depths of the shocked and unshocked CSM components ahead of the diffusion front, respectively, and $m_{\rm SC,diff}$ is the corresponding diffusion mass in the shocked CSM. Including unshocked CSM mass $m_{\rm UC,diff}\sim4\pi\rho_{\rm CSM}(r_{\rm out}^3-r_{\rm sh}^3)/3$, the total diffusion mass can be estimated as
\begin{equation}
\begin{split}
    m_{\rm diff}&\sim m_{\rm SC,diff}+m_{\rm UC,diff}\\
    &=\frac{M_{\rm CSM}(r_{\rm out}^3+2r_{\rm sh}^3-3r_{\rm SBO}r_{\rm sh}^2)}{r_{\rm out}^3-r_{\rm in}^3}.
\end{split}
\end{equation}
{We assume that shock heating and cooling emission uniformly heat all material exterior to the diffusion front.} Then, the internal energy deposited by shock heating and cooling in the material outside the diffusion front is given by
\begin{equation}
    E_{\rm int}\sim \frac{m_{\rm diff}v_{\rm sh}^2}{2},
\end{equation}
where adiabatic cooling can be neglected because the post-shock density remains nearly constant during this phase. The release of this internal energy over a timescale $t$ gives rise to a bolometric luminosity of
\begin{equation}
\label{equ:LightcurveSBOPhase}
    L_{\rm sou}(t_{0}<t\leq2t_0)\sim\frac{E_{\rm int}}{t}\sim\frac{m_{\rm diff}v_{\rm sh}^2}{2t}.
\end{equation}

In the thin-shell limit for the CSM, the diffusion mass is expected to satisfy
$m_{\rm diff}\sim M_{\rm CSM}(r_{\rm out}-r_{\rm SBO})/(r_{\rm out}-r_{\rm in})\sim m_{\rm SBO}$, which remains nearly constant. Therefore, during the shock emergence phase, the source lightcurve can be approximated as
\begin{equation}
\label{equ:LightcurveShockEmergencePhase}
    L_{\rm sou}^{\rm thin}(t_0<t\leq2t_0)\sim \frac{L_{\rm SBO}t_0}{t}
\end{equation}

(3) {\em Cooling Phase.} After $t>2t_0$, we assume that all swept-up CSM material is concentrated in a thin shell. The diffusion front is then determined by
\begin{equation}
    \tau_{\rm diff}\approx \frac{\kappa m_{\rm SC,diff}}{4\pi r_{\rm sh}^2}\sim\frac{c}{v_{\rm sh}},
\end{equation}
in which case the diffusion mass is entirely contributed by the shocked CSM shell, i.e.,
\begin{equation}
\label{equ:DiffusionMassCoolingPhase}
    m_{\rm diff} = m_{\rm SC,diff}=\frac{3M_{\rm CSM}r_{\rm sh}^2(r_{\rm out}-r_{\rm SBO})}{r_{\rm out}^3-r_{\rm in}^3}.
\end{equation}
Then, the internal energy can be expressed as
\begin{equation}
\label{equ:InternalEnergyCoolingPhase}
    {E_{\rm int}\sim \frac{m_{\rm diff}v_{\rm sh}^2}{2}\left(\frac{r_{\rm out}}{r_{\rm sh}}\right),}
\end{equation}
where the factor {$(r_{\rm out}/r_{\rm sh})$}  accounts for the adiabatic cooling due to the {spherical homologous} expansion of the shocked shell. One can derive the bolometric luminosity as
\begin{equation}
\label{equ:LightcurveCoolingPhase}
\begin{split}
    &L_{\rm sou}(t>2t_0)\sim \\
    &\begin{cases}
       \dfrac{E_{\rm int}(t)}{t}, & {\rm if\ }2t_0<t\leq t_{\rm diff,c}, \\
       \dfrac{E_{\rm int}(t_{\rm diff,c})}{t_{\rm diff,c}}e^{-(t^2/t_{\rm diff,c}^2-1)/2}, & {\rm if\ }t>t_{\rm diff,c},
       \end{cases}
\end{split}
\end{equation}
where 
\begin{equation}
    t_{\rm diff,c}\approx2t_0+\frac{1}{v_{\rm sh}}\left[\sqrt{\frac{r_{\rm out}^3-r_{\rm in}^3}{3(r_{\rm out}-r_{\rm SBO})}}-r_{\rm out}\right],
\end{equation}
is the characteristic diffusion timescale for the diffusion front to traverse the entire shocked CSM (i.e., $m_{\rm diff}\sim M_{\rm CSM}$), after which the lightcurve decays exponentially \citep[e.g.,][]{Piro2021,Zhu2025}. 

Under the thin-shell approximation for the CSM, $m_{\rm diff}$ in Equation (\ref{equ:DiffusionMassCoolingPhase}) reduces to $m_{\rm diff}\sim m_{\rm SBO}(r_{\rm sh}/r_{\rm out})^2$, while $E_{\rm int}$ in Equation (\ref{equ:InternalEnergyCoolingPhase}) can be further written as $E_{\rm int}\sim E_{\rm SBO}(r_{\rm sh}/r_{\rm out})$. By defining the transition timescale $t_{\rm sph}\approx r_{\rm out}/v_{\rm sh}+2t_0$, corresponding to the time when $r_{\rm sh}$ expands to twice $r_{\rm out}$, i.e., the transition from the planar phase to the spherical phase, we find that $E_{\rm int}\sim E_{\rm SBO}$ for $t\ll t_{\rm sph}$, while $E_{\rm int}\propto t$ for $t\gg t_{\rm sph}$. By combining Equations (\ref{equ:LightcurveSBOPhase}), (\ref{equ:LightcurveShockEmergencePhase}), and (\ref{equ:LightcurveCoolingPhase}), we obtain an empirical expression for the bolometric lightcurve as
\begin{equation}
\begin{split}
    &L_{\rm sou}^{\rm thin}\sim L_{\rm SBO}\\
    &\times\begin{cases}
       e^{1-t_0/t}, & {\rm if\ }t<t_{\rm 0}, \\
       \dfrac{t_0}{t}, & {\rm if\ }t_0<t\leq t_{\rm sph}, \\
       {\dfrac{t_0}{t_{\rm sph}}}, & {\rm if\ }t_{\rm sph}<t\leq t_{\rm diff,c},\\
       {\dfrac{t_{\rm 0}}{t_{\rm sph}}e^{-(t^2/t_{\rm diff,c}^2-1)/2}}, & {\rm if\ }t>t_{\rm diff,c},
       \end{cases}
\end{split}
\end{equation}
if $t_{\rm sph}<t_{\rm diff,c}$, or
\begin{equation}
\begin{split}
    &L_{\rm sou}^{\rm thin}\sim L_{\rm SBO}\\
    &\times\begin{cases}
       e^{1-t_0/t}, & {\rm if\ }t<t_{\rm 0}, \\
       \dfrac{t_0}{t}, & {\rm if\ }t_0<t\leq t_{\rm diff,c}, \\
       \dfrac{t_0}{t_{\rm diff,c}}e^{-(t^2/t_{\rm diff,c}^2-1)/2}, & {\rm if\ }t>t_{\rm diff,c},
       \end{cases}
\end{split}
\end{equation}
if $t_{\rm sph}\geq t_{\rm diff,c}$.

During the SBO phase ($t<t_0$), we freeze the source color temperature as the value at the breakout time, i.e., $T_{\rm sou,col}(t<t_0)=T_{\rm col,SBO}$. For the post-breakout phase ($t\geq t_0$), we calculate the source equilibrium blackbody temperature $T_{\rm sou,BB}$ by modeling the photon energy density at the diffusion front based on the source luminosity, following Equation (\ref{equ:BBTemperature}) as
\begin{equation}
    \epsilon\sim\frac{L_{\rm sou}\tau_{\rm diff}}{4\pi r_{\rm sh}^2c(1-e^{-\tau_{\rm diff}})}.
\end{equation}
After considering Compton equilibrium, the source color temperature $T_{\rm sou,col}(t\geq t_0)$ can be calculated using Equations (\ref{equ:eta}) and (\ref{equ:ColorTemperature}), with $t_{\rm diff}=t$ for the shocked CSM shell at the diffusion front and the shocked CSM density given by
\begin{equation} 
\rho_{\rm sh}\approx
    \begin{cases} 7\rho_{\rm CSM}, & {\rm if}\ t\leq2t_0,\\ 
    7\rho_{\rm CSM}{(r_{\rm out}/r_{\rm sh})^3}, & {\rm if}\ t>2t_0. 
    \end{cases} 
\end{equation}

\subsection{Observed Lightcurve and Emergent Spectra with Light-travel Effects}

Due to the light-travel time effect, photons emitted from different surface elements arrive at the observer at different times. The emission time $t$ corresponding to an observed time $t_{\rm obs}$ is therefore related to the viewing angle $\theta$ (i.e., the angle between the line of sight and the direction of the emitting surface element) by
\begin{equation}
    t=t_{\rm obs} - \frac{\max(r_{\rm out},r_{\rm sh})}{c}(1-\cos\theta).
\end{equation}
Since the specific intensity satisfies $I\propto L_{\rm sou}(t)$, the observed luminosity is obtained by integrating the emission over the equal-arrival-time surface:
\begin{equation}
\begin{split}
\label{equ:ObservedLuminosity}
    L_{\rm obs}(t_{\rm obs})&\propto\int I\cos\theta dA \\
    &\propto\int_0^{\pi/2}L_{\rm sou}[t(\theta,t_{\rm obs})]\sin\theta\cos\theta d\theta,
\end{split}
\end{equation}
where $dA$ is the spherical area element.

(1) We first consider the case where the shock front is still inside the CSM shell, i.e., $r_{\rm sh}\leq r_{\rm out}$ (or equivalently $t<2t_0$). In this regime, the emitting radius is fixed around $r_{\rm out}$, and the emission time is therefore related to the viewing angle by 
\begin{equation}
\label{equ:ObservedTimescaleInCSM}
    t = t_{\rm obs} -\frac{r_{\rm out}}{c}(1-\cos\theta).
\end{equation}
For a given $t_{\rm obs}$, the earliest and latest emission times contributing to the observed luminosity correspond to the edge ($\theta=\pi/2$) and the line-of-sight direction ($\theta=0$), respectively, which correspond to $t_{\rm min,in}=\max[0,\min(2t_0,t_{\rm obs}-r_{\rm out}/c)]$ and $t_{\rm max,in}=\min(2t_0,t_{\rm obs})$. Using the relation $\sin\theta d\theta=-(c/r_{\rm out})dt$ from Equation (\ref{equ:ObservedTimescaleInCSM}), the angular integration in Equation (\ref{equ:ObservedLuminosity}) can be transformed into an integral over the emission time, yielding the observed luminosity in this regime,
\begin{equation}
\label{equ:ObservedLuminosityIn}
    L_{\rm obs,in}(t_{\rm obs}) = \frac{2c}{r_{\rm out}}\int_{t_{\rm min,in}}^{t_{\rm max,in}}L_{\rm sou}(t)\left(1-\frac{t_{\rm obs}-t}{r_{\rm out}/c}\right)dt,
\end{equation}
where the factor $2$ {is introduced to normalize the angular integral over the visible hemisphere, since $\int_0^{\pi/2}\sin\theta\cos\theta d\theta=1/2$.} 

(2) In the regime after the shock emergence phase, i.e., $r_{\rm sh}=r_{\rm out}+v_{\rm sh}(t-2t_{\rm 0})$, the emission time is related to the observed time by
\begin{equation}
    t = t_{\rm obs}-\frac{r_{\rm out}+v_{\rm sh}(t-2t_0)}{c}(1-\cos\theta),
\end{equation}
{which can be rearranged to give}
\begin{equation}
    t(\theta) = \frac{ct_{\rm obs} - (r_{\rm out}-2v_{\rm sh}t_0)(1-\cos\theta)}{c+v_{\rm sh}(1-\cos\theta)}.
\end{equation}
The angular integration element can then be expressed as
\begin{equation}
    \sin\theta d\theta = -\frac{c[r_{\rm out}+v_{\rm sh}(t_{\rm obs}-2t_0)]}{[r_{\rm out}+v_{\rm sh}(t-2t_0)]^2}dt=-\frac{cr_{\rm sh}(t_{\rm obs})}{r_{\rm sh}(t)^2}dt.
\end{equation}
In this regime, for a given $t_{\rm obs}$, the earliest emission times to the edge is $t_{\rm min,out}=\max\{2t_0,[ct_{\rm obs}-(r_{\rm out}-2v_{\rm sh}t_0)]/(c+v_{\rm sh})\}$, while the latest emission time in the line-of-sight direction is $t_{\rm max,out}=\max(2t_0 , t_{\rm obs})$. Thus, the observed luminosity in Equation (\ref{equ:ObservedLuminosity}) can be given by
\begin{equation}
\begin{split}
\label{equ:ObservedLuminosityOut}
    L_{\rm obs,out}&(t_{\rm obs})  = 2cr_{\rm sh}(t_{\rm obs})\\
    &\times\int_{t_{\rm min,out}}^{t_{\rm max,out}}\frac{L_{\rm sou}(t)}{r_{\rm sh}(t)^2}\left[1-\frac{t_{\rm obs}-t}{r_{\rm sh}(t)/c}\right]dt.
\end{split}
\end{equation}
The total observed luminosity is thus $L_{\rm obs} = L_{\rm obs,in}+L_{\rm obs,out}$.

The observed spectral luminosities $L_{\nu,\rm obs,in}$ and $L_{\nu,\rm obs,out}$ can be obtained following the same formalism as Equations (\ref{equ:ObservedLuminosityIn}) and (\ref{equ:ObservedLuminosityOut}), by replacing the source bolometric luminosity $L_{\rm sou}(t)$ with the source spectral luminosity
\begin{equation}
    L_{\nu\rm ,sou}(\nu,t) = L_{\rm sou}(t)\frac{\mathcal{B}_\nu[\nu,T_{\rm sou,col}(t)]}{\int_0^{+\infty}\mathcal{B}_\nu[\nu,T_{\rm sou,col}(t)]d\nu},
\end{equation}
where $\mathcal{B}_\nu$ represents the spectral shape, which is taken to be the Planck function for a thermalized blackbody spectrum when the LTE condition $\eta\leq1$ is satisfied (typically for $v_{\rm sh}\lesssim0.05\,c$), and as the Wien spectrum for a Comptonized spectrum when the non-LTE condition $\eta>1$ is applied (typically for $v_{\rm sh}\gtrsim0.05\,c$). To describe this transition, we phenomenologically introduce the following spectral form:
\begin{equation}
    \mathcal{B}_\nu\propto\frac{\nu^3}{\exp(h\nu/k_BT_{\rm sou,col})-[\max(1,\eta)]^{-1}},
\end{equation}
where $h$ is the Planck constant. The total observed spectral luminosity is then given by $L_{\nu,\rm obs}=L_{\nu,\rm obs,in} + L_{\nu,\rm obs,out}$, which generally represents a multi-temperature thermal spectrum because the observed emission is contributed by photons emitted at different times with different source temperatures due to the light-travel time effect. The observed X-ray lightcurve can be expressed as $L_{\rm X}({t_{\rm obs}})=\int L_{\nu,\rm obs}(t_{\rm obs})d\nu$, integrated over the bandpass of the X-ray telescopes.

\subsection{Results}

\subsubsection{Observed Lightcurve, Temperature Evolution, and Observed Spectra} \label{sec:Spectra}

\begin{figure}[t!]
\centering
\includegraphics[width=1\linewidth, trim = 41 25 126 58, clip]{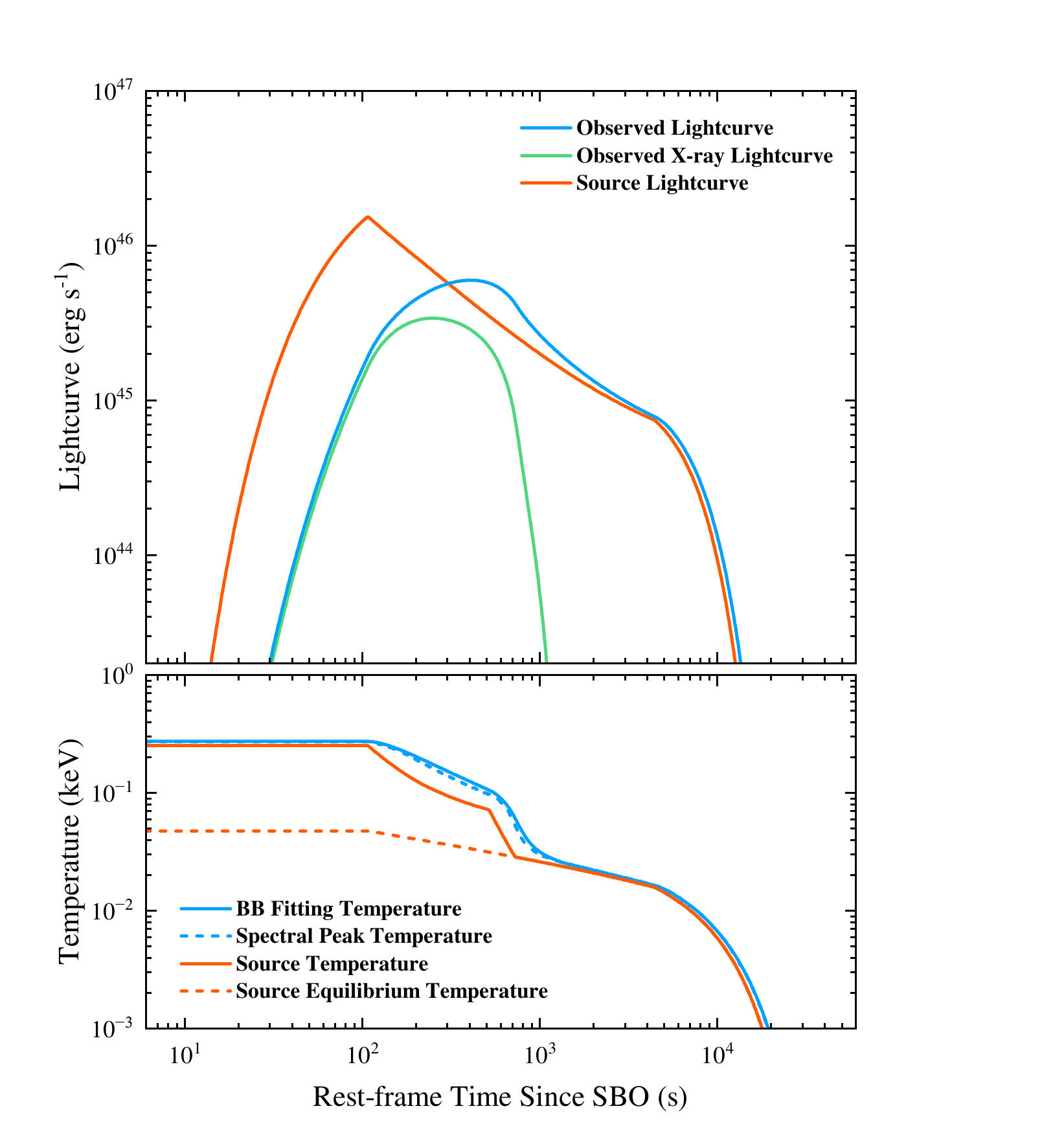}
\caption{Lightcurves (top panel) and temperature evolution (bottom panel) for an example model. In the top panel, the blue, green, and orange lines denote the observed lightcurve including the light-travel effect, the observed X-ray lightcurve in the WXT bandpass, and the source lightcurve, respectively. The bottom panel shows the observed temperature obtained from blackbody fits, the observed spectral peak temperature, the source color temperature, and the source equilibrium blackbody temperature, represented by the blue solid, blue dashed, orange solid, and orange dashed lines, respectively. }
\label{fig:Example_Lightcurve}
\end{figure}

\begin{figure}[t!]
\centering
\includegraphics[width=1\linewidth, trim = 38 34 127 3, clip]{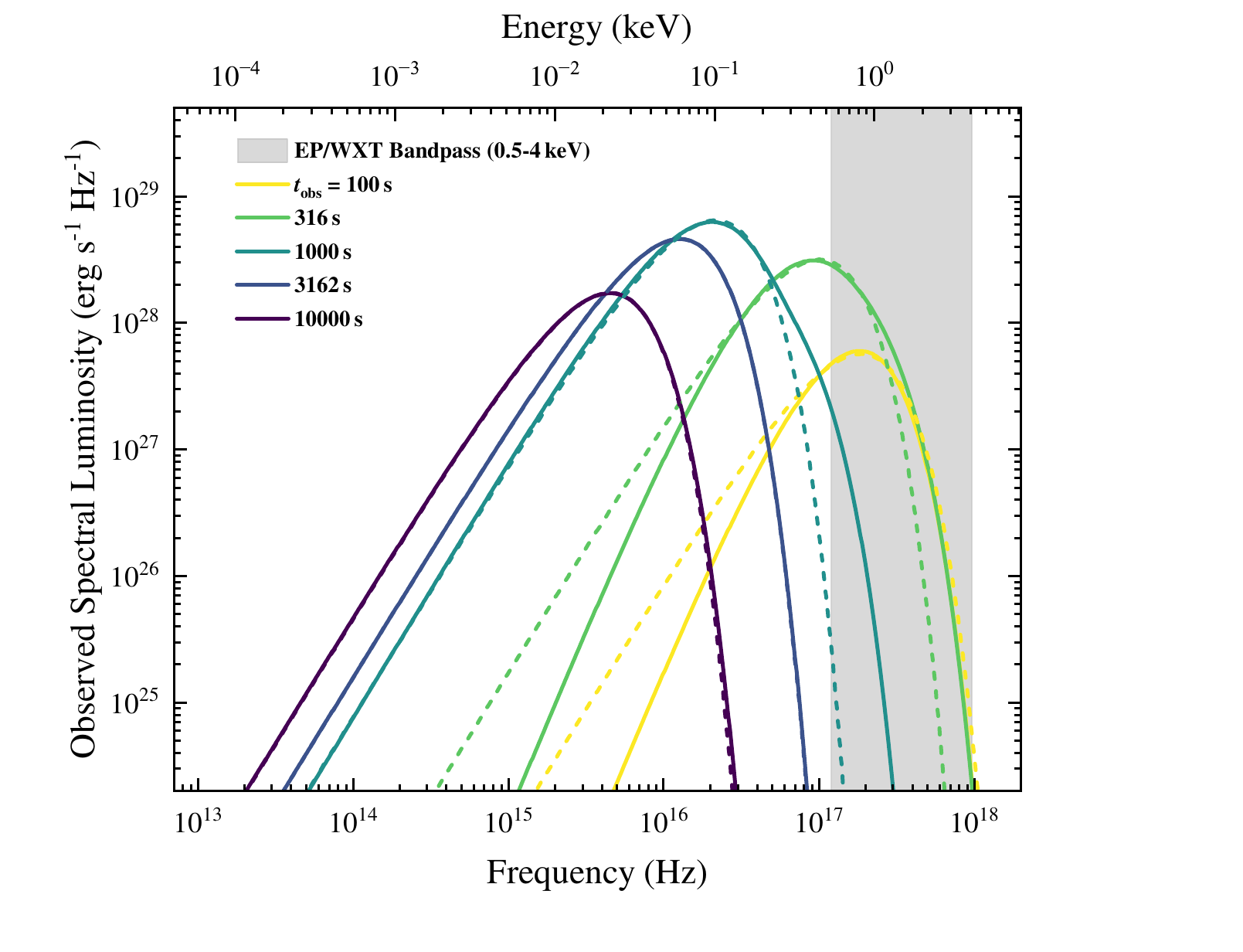}
\caption{Multi-epoch observed spectra of the SBO and cooling phases for an example model. The solid colored lines show the simulated observed spectra from $t_{\rm obs}=10^2\,{\rm s}$ to $10^4\,{\rm s}$, while the dashed lines represent the corresponding blackbody fits at each epoch. The gray shaded region marks the WXT bandpass ($\sim0.5-4\,{\rm keV}$).   }
\label{fig:Example_Spectrum}
\end{figure}

As an illustrative example, we consider a model with parameters $r_{\rm out}=2\times10^{13}\,{\rm cm}$, $v_{\rm sh}=0.15\,c$, $\chi=0.1$, and $M_{\rm CSM}=3.2\times10^{-4}\,M_\odot$. The observed lightcurves and temperature evolution of the SBO and cooling phase are shown in Figure \ref{fig:Example_Lightcurve}. Compared with the source lightcurve, the observed lightcurve is affected by the light-travel effect, which reduces the peak luminosity, smooths and broadens the profile around the peak, and leads to a longer observed peak timescale. When the observed time is significantly longer than the light-travel time, i.e., {$t_{\rm obs}\gg t_{\rm lt}\sim r_{\rm out}/c\approx700\,r_{\rm out,13.3}\,{\rm s}$}, the observed lightcurve gradually approaches the source lightcurve, indicating that the light-travel-time effect is negligible.

As shown in Figure \ref{fig:Example_Lightcurve}, the source color temperature exceeds the source equilibrium blackbody temperature by nearly an order of magnitude during the early phase because of the non-LTE. The enhanced color temperature shifts the spectral peak toward the soft X-ray band, making the SBO emission detectable by soft X-ray instruments. After $10^3\,{\rm s}$, the two temperatures become consistent, indicating that LTE has been established around the diffusion front. Furthermore, Figure \ref{fig:Example_Spectrum} presents the observed spectra at representative epochs between $10^2$ and $10^4\,{\rm s}$. When the observed time is shorter or comparable to the light-travel time (i.e., $t_{\rm obs}\lesssim t_{\rm lt}$), the observed spectra deviate from a blackbody because the light-travel effect mixes emissions from regions with different temperatures. Nevertheless, around the spectral peak, the observed spectra can still be well reproduced by a blackbody fit. At $t_{\rm obs}\gg t_{\rm lt}$, the temperature gradient across the emitting surface becomes much smaller, and the spectra become closer to those of a single-temperature blackbody. In Figure \ref{fig:Example_Lightcurve}, we also show the evolution of the observed temperature derived from blackbody fitting and the temperature estimated from the spectral peak, which show good agreement with each other. For $t_{\rm obs}\lesssim t_{\rm lt}$, both temperatures show a temporal delay and smoothing compared with the source color temperature. However, when $t_{\rm obs}\gg t_{\rm lt}$, they closely follow the source color temperature, with only negligible differences.

\subsubsection{Parameter Dependence} \label{sec:ParameterDependence}

\begin{figure*}[t!]
\centering
\includegraphics[width=0.447\linewidth, trim = 41 25 126 58, clip]{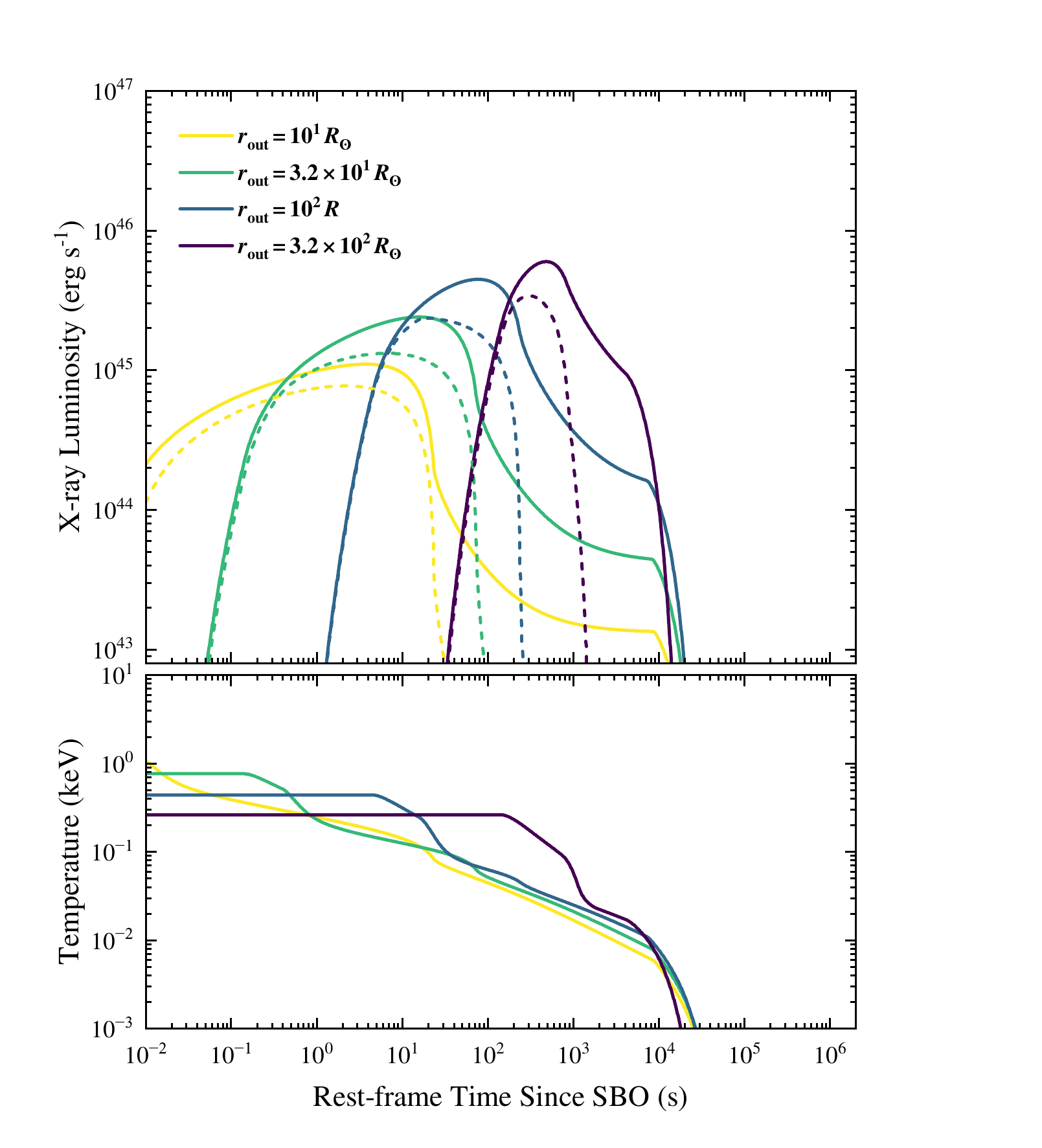}
\includegraphics[width=0.447\linewidth, trim = 41 25 126 58, clip]{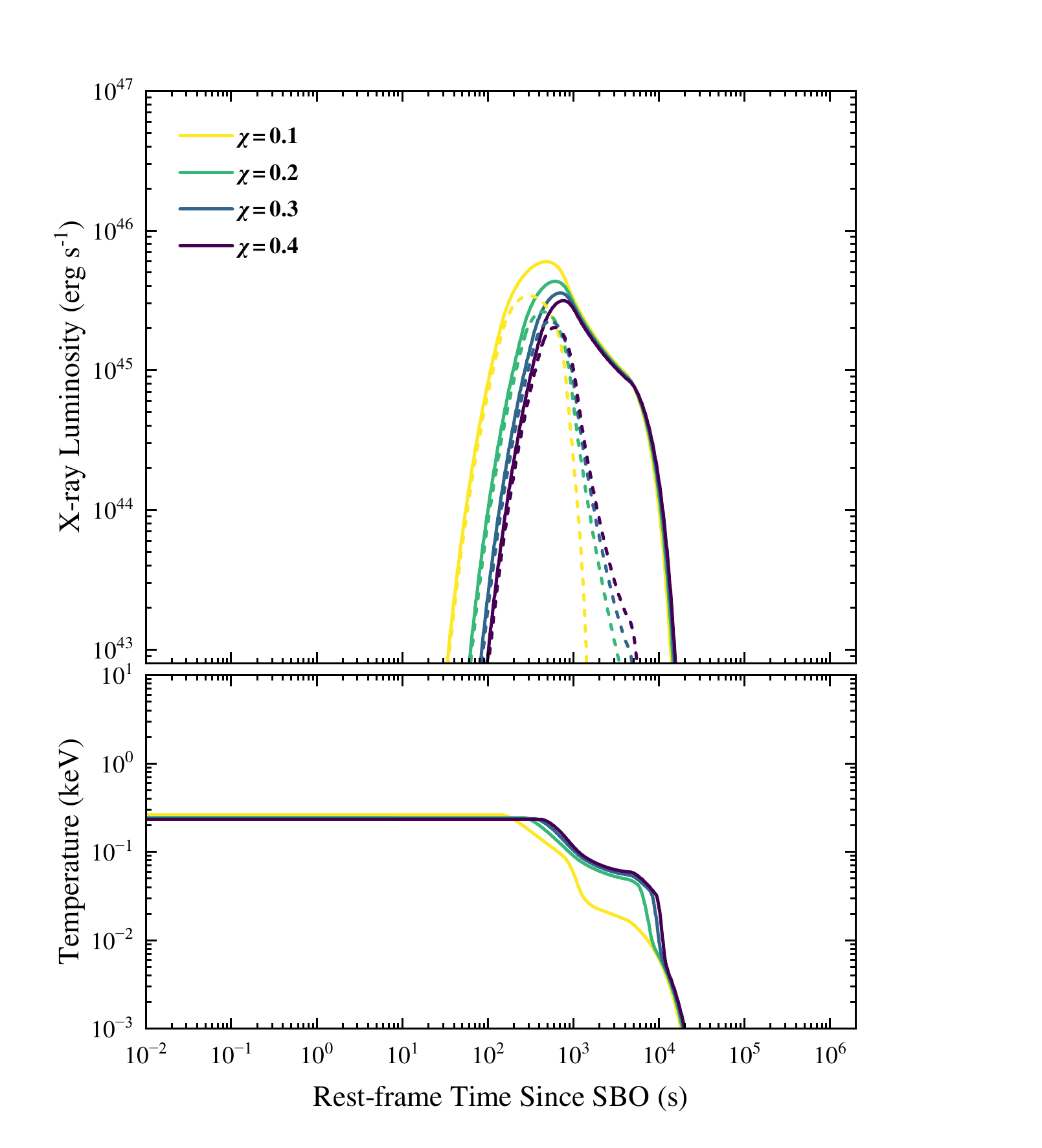}
\includegraphics[width=0.447\linewidth, trim = 41 25 126 58, clip]{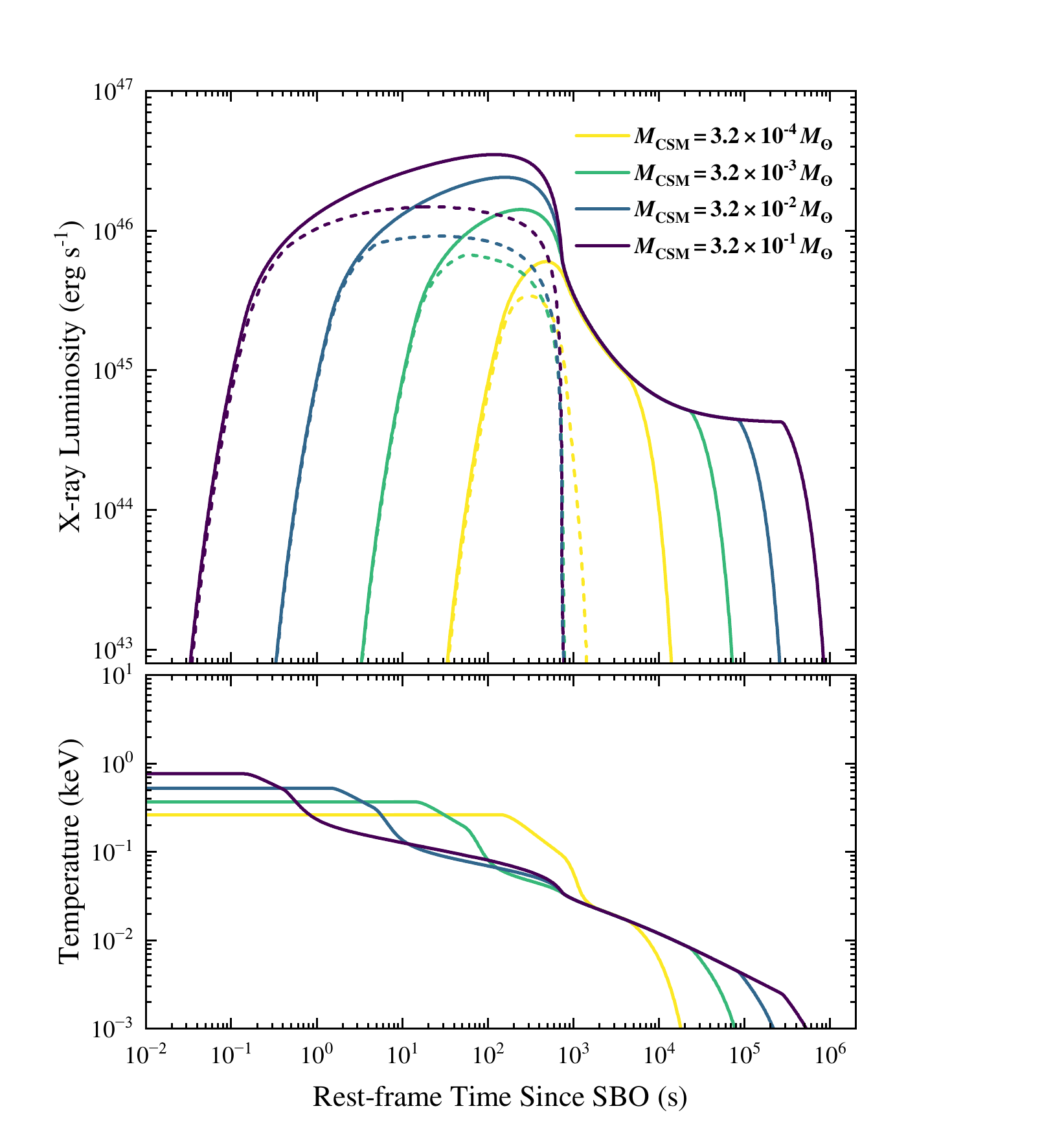}
\includegraphics[width=0.447\linewidth, trim = 41 25 126 58, clip]{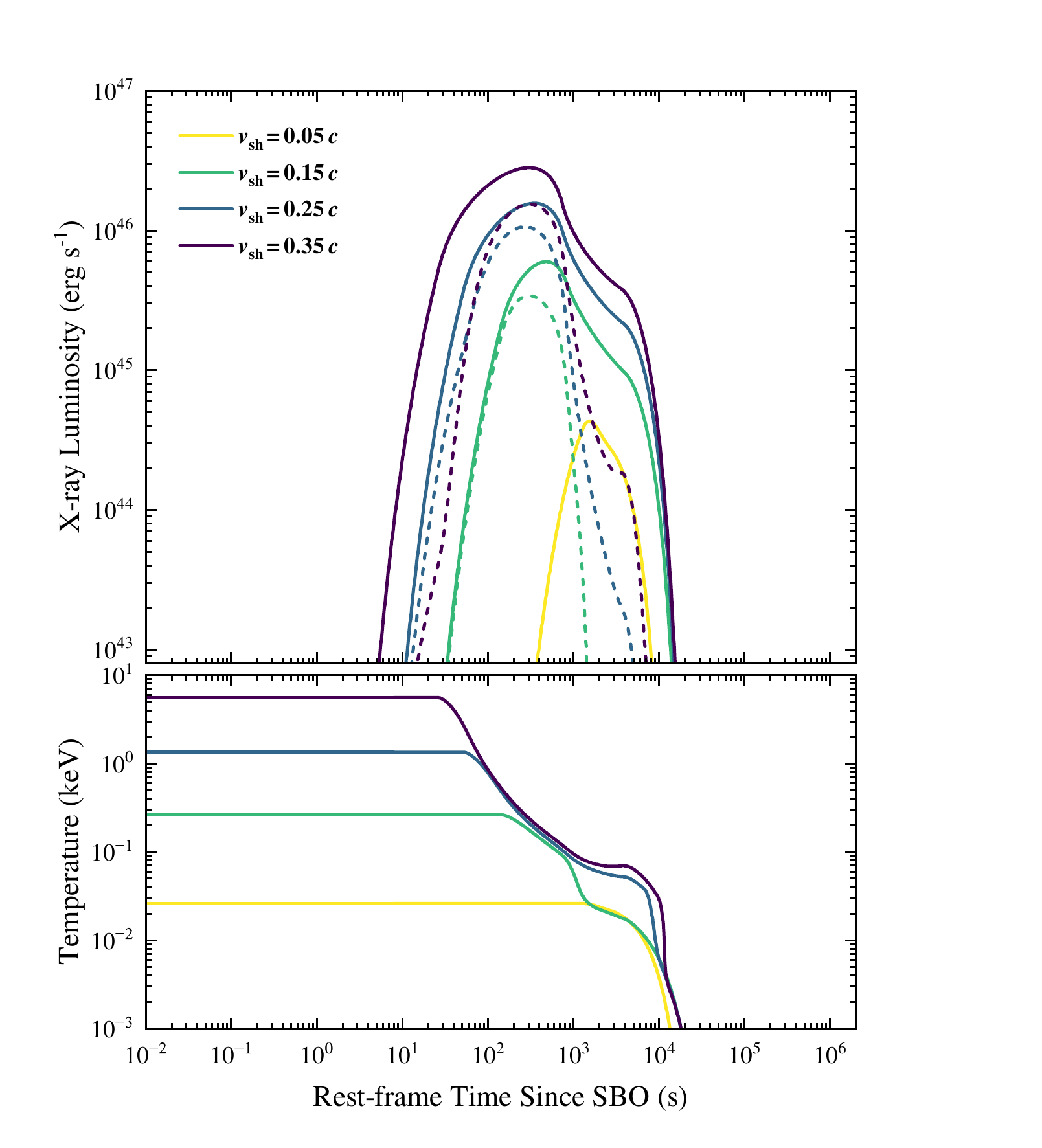}
\caption{Simulated observed lightcurves and spectral peak temperature evolution of SN SBO and subsequent cooling emission in dense CSM shell for different $r_{\rm out}$ (top left), $\chi$ (top right), $M_{\rm CSM}$ (bottom left), and $v_{\rm sh}$ (bottom right). In each panel, the upper sub-panel presents the observed lightcurves in both the bolometric band (solid lines) and soft X-ray band (dashed lines; $0.5-4\,{\rm keV}$), while the lower sub-panel shows the evolution of the observed spectral peak temperature. Each panel shows four choices of the corresponding parameter, including the fiducial values of $r_{\rm out}=3.2\times10^{2}\,R_\odot$, $\chi=0.1$, $M_{\rm CSM}=3.2\times10^{-4}\,M_\odot$, and $v_{\rm sh}=0.15\,c$.  }
\label{fig:Parameter_Dependence}
\end{figure*}

We consider a range of CSM parameters and shock velocities to cover the diversity of observed lightcurves and spectral peak temperature evolutions, and to investigate the dependence of these observables on the model parameters. Specifically, we vary the CSM outer radius, dimensionless shell thickness, CSM mass, and shock velocity between $10^1\,R_\odot\leq r_{\rm out}\leq3.2\times10^2\,R_\odot$, $0.1\leq\chi\leq0.4$, $3.2\times10^{-4}M_\odot\leq M_{\rm CSM}\leq3.2\times10^{-1}\,M_\odot$, and $0.05\,c\leq v_{\rm sh}\leq0.35\,c$, respectively. Adopting the fiducial parameters of $r_{\rm out}=3.2\times10^{2}\,R_\odot$, $\chi=0.1$, $M_{\rm CSM}=3.2\times10^{-4}\,M_\odot$, and $v_{\rm sh}=0.15\,c$, we show the bolometric and X-ray lightcurves, as well as the spectral peak temperature evolution, in Figure \ref{fig:Parameter_Dependence} by varying each parameter over four different choices. These models produce soft X-ray transients with peak X-ray luminosities ranging from $\sim10^{45}-10^{46}\,{\rm erg}\,{\rm s}^{-1}$. The X-ray emission typically peaks within a few times $10^{3}\,{\rm s}$ after the explosion and lasts for less than a few times $10^{6}\,{\rm s}$. The spectral peak temperature at SBO is always within the soft X-ray band for most considered models.

{\em Dependence on Outer Radius.} As the $r_{\rm out}$ increases, both the bolometric and soft X-ray luminosities become higher, primarily due to the larger emitting surface area at SBO. Meanwhile, the increased light-travel effects in the more extended CSM shell become more significant, leading to a delayed lightcurve peak and a longer duration of the soft X-ray emission. In contrast, the maximum temperature, corresponding to the temperature at the SBO phase, decreases with increasing $r_{\rm out}$. Nevertheless, the maximum temperature remains within the soft X-ray band for all considered $r_{\rm out}$ values. For smaller $r_{\rm out}$, the planar phase becomes shorter, allowing the emission from the subsequent spherical phase to emerge at earlier times.

{\em Dependence on Dimensionless Shell Thickness.} The effect of $\chi$ on both the lightcurves and temperature evolution is relatively minor. For fixed $M_{\rm CSM}$, $r_{\rm out}$, and $v_{\rm sh}$, a larger $\chi$ corresponds to a wider CSM shell with a lower density at the SBO radius. As a result, the SBO luminosity is slightly reduced because it is positively correlated with the upstream unshocked CSM density.

{\em Dependence on Circumstellar Mass.} Increasing $M_{\rm CSM}$ has only a minor effect on the duration of the soft X-ray SBO emission, while the peak X-ray luminosity increases moderately due to the higher unshocked CSM density at the SBO. A larger $M_{\rm CSM}$ significantly enhances the late-time UV-to-optical emission and increases its duration, potentially producing bright SN-like emission for $M_{\rm CSM}\gg10^{-2}\,M_\odot$ \citep[e.g.,][]{Chevalier2011,Chatzopoulos2012,Khatami2024,Zhang2026}.

{\em Dependence on Shock Velocity.} Once the CSM properties are fixed, $v_{\rm sh}$ is primarily determined by the ejecta kinetic energy $E_{\rm kin}$ and ejecta mass $M_{\rm ej}$ (see Equations \ref{equ:ShockVelocity1} and \ref{equ:ShockVelocity2}). A larger $E_{\rm kin}$ or a lower $M_{\rm ej}$ naturally leads to a higher $v_{\rm sh}$ and thus more powerful SBO and cooling emission. As discussed in \cite{Nakar2010} and \cite{Katz2010}, producing soft X-ray SBO emission generally requires shock velocities above $v_{\rm sh}\gtrsim0.1\,c$, where the shock radiation may deviate from blackbody thermal emission due to non-LTE. Consistent with this expectation, we find that models with $v_{\rm sh}\sim0.05\,c$ fail to produce bright soft X-ray SBO and cooling emission. For $v_{\rm sh}\gtrsim0.15\,c$, the durations of the soft X-ray SBO and cooling emission become nearly insensitive to $v_{\rm sh}$, while their peak luminosities continue to increase with increasing $v_{\rm sh}$ due to the stronger shock power.

\section{Case Studies}

Following the model proposed in Section \ref{sec:Model}, we then adopt a Markov Chain Monte Carlo method with the \texttt{emcee} package \citep{ForemanMackey2013} to explain the X-ray lightcurve and temperature evolution of potential SBO events, including EP260321a \citep{Yuan2026}, XRF\,060218 \citep{Campana2006}, and XRO\,080109 \citep{Soderberg2008}. Here, we model the observed temperature using the spectral peak temperature. There are {seven} free parameters for the SBO and cooling emission, including the outer radius $r_{\rm out}$, dimensionless shell thickness $\chi$, CSM mass $M_{\rm CSM}$, shock velocity $v_{\rm sh}$, time of SBO relative to the first observed data point $t_{\rm first}$, {and fractional model-uncertainty parameters in the X-ray luminosity and color temperature $f_{\rm L}$ and $f_{\rm T}$}, respectively. {Details of the fitting procedure, prior distributions, and representative posterior distributions are provided in Appendix \ref{app_sec:MCMC}.} Our subsequent fitting results reveal significant diversity in the inferred CSM properties among the three X-ray SBO events, indicating diverse pre-SN mass-loss histories of stripped-envelope progenitors. 

\subsection{EP260321a}

Recently, the EP mission reported the discovery of a soft X-ray SBO, EP260321a, at a redshift of $z=0.0344$ \citep{Yuan2026}, which is associated with a broad-lined Type Ic SN, SN\,2026gzf \citep{Yuan2026,Chen2026,MartinCarrillo2026,Rastinejad2026,OConnor2026,Wen2026}. Its X-ray emission peaks at $T_0+(690\pm60)\,{\rm s}$ with a peak luminosity of $(1.0\pm0.3)\times10^{45}\,{\rm erg}\,{\rm s}^{-1}$. The total radiated energy in the observed $0.3-10\,{\rm keV}$ X-ray band is $(8.7\pm0.6)\times10^{47}\,{\rm erg}$. The X-ray spectra of EP260321a obtained by WXT (from $T_0$ to $T_0+2200\,{\rm s}$) and FXT (from $T_0+1065\,{\rm s}$ to $T_0+2359\,{\rm s}$) can be well described by an absorbed blackbody model, with blackbody temperatures of $124^{+7}_{-6}\,{\rm eV}$ and $112.7^{+2.1}_{-2.1}\,{\rm eV}$, respectively. 

As discussed in Section \ref{sec:Spectra}, the observed spectra are expected to be affected by the light-travel effect, which leads to the integration of emission from different radii and hence a multi-temperature thermal spectrum rather than a single-temperature blackbody. However, given the limited energy coverage of EP, the observed spectra around the spectral peak can still be reasonably approximated by a blackbody function, resulting in an apparently good blackbody fit. Therefore, a blackbody-like time-resolved spectrum in the soft X-ray band does not necessarily indicate a truly single-temperature emission component, as the limited bandpass may hide deviations from a Planck spectrum caused by the Wien-like spectral shape and the underlying multi-temperature structure.

\cite{Yuan2026} suggested that the SBO and cooling of the SN in a dense CSM shell can simultaneously fit EP260321a's X-ray lightcurves observed by WXT and FXT in the $0.4-2\,{\rm keV}$ band and the observed temperature evolution. The posterior distributions are {$r_{\rm out}=322^{+30}_{-30}\,R_\odot$, $v_{\rm sh}=0.129^{+0.007}_{-0.005}\,c$, $\chi=0.334^{+0.040}_{-0.038}$, $M_{\rm CSM}=9.0^{+2.8}_{-1.3}\times10^{-5}\,M_\odot$, $t_{\rm first}=1089^{+214}_{-203}\,{\rm s}$, $f_L = 0.058^{+0.051}_{-0.033}$, and $f_T = 0.244^{+0.091}_{-0.060}$, respectively. The posterior constraints imply optical depths of $\tau_{\rm CSM}=8.08^{+0.38}_{-0.30}$ for the entire CSM shell and $\tau_{\rm SBO}=7.76^{+0.33}_{-0.38}$ at breakout, as well as an inner radius of $r_{\rm in}=215^{+31}_{-29}\,R_\odot$ and a SBO radius of $r_{\rm SBO}=218^{+29}_{-25}\,R_\odot$, suggesting that the SBO occurs close to the inner boundary of the CSM shell. The thermal energy deposited by the shock can be estimated as $E_{\rm th}\sim M_{\rm CSM}v_{\rm sh}^2/2\sim1.3\times10^{48}\,{\rm erg}$, comparable to the observed X-ray radiated energy of $\sim9\times10^{47}\,{\rm erg}$ from EP260321a.  }

\subsection{XRF\,060218}

\begin{figure}[t!]
\centering
\includegraphics[width=1\linewidth, trim = 41 25 126 58, clip]{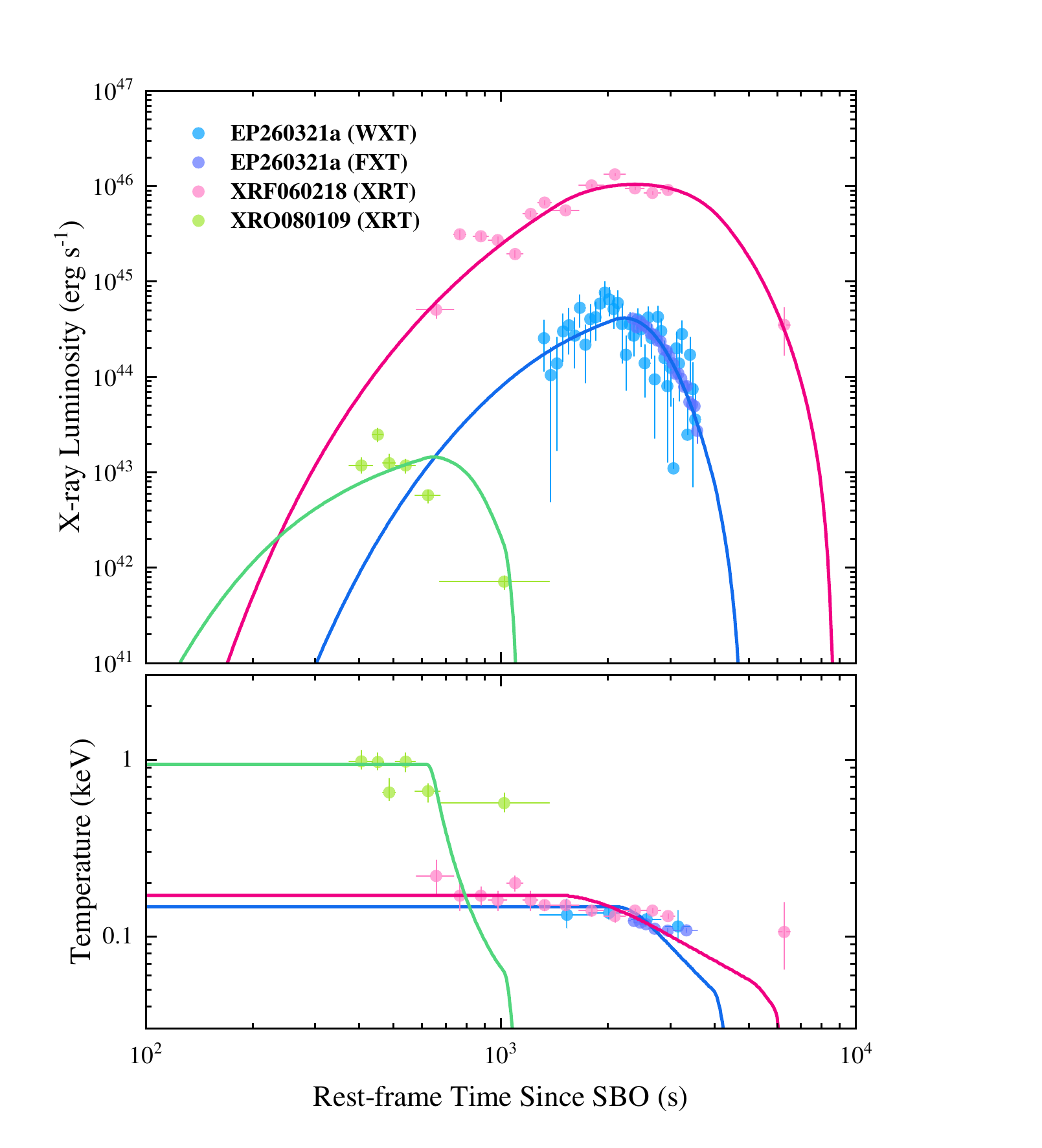}
\caption{X-ray lightcurve (top panel) and temperature evolution (bottom panel) of EP260321a, the thermal component in XRF\,060218, and XRO\,080109. The blue, {pink, and green} points represent the data of EP260321a, XRF\,060218, and XRO\,080109, respectively. The final {pink} point's large error and deviation from the fit for XRF\,060218 likely arise from incomplete non-thermal component subtraction and a low photon count. The solid lines show the corresponding best-fit models.  }
\label{fig:Cases}
\end{figure}

Confirmed to be connected with the broad-lined Type Ic SN SN\,2006aj, XRF\,060218 was first detected by the {\em Swift}/BAT as a soft X-ray flash with non-thermal prompt emission at a redshift of $z=0.033$ \citep{Campana2006,Pian2006,Mazzali2006}. Follow-up observations with {\em Swift}/XRT revealed an additional thermal component in the soft X-ray ($0.3-10\,{\rm keV}$) spectra, with a peak luminosity of $\sim10^{46}\,{\rm erg\,s^{-1}}$ and a duration of up to $\sim10^4\,{\rm s}$. Here, we adopt the spectral parameters of the thermal and non-thermal components from \citet{Kaneko2007} and refit the time-resolved XRT spectra to derive the X-ray luminosity of the thermal component in the {$0.3-10\,{\rm keV}$} band. For the last time interval, we fitted the spectrum with a blackbody plus a power-law model. The blackbody X-ray component showed a mildly decreasing temperature, with the observed temperature of $\sim0.17\,{\rm keV}$. 

As discussed regarding the blackbody-like time-resolved spectra of EP260321a, the thermal component of XRF\,060218, although well described by a blackbody model, may also result from the limited spectral coverage and the superposition of multi-temperature emission components within the observed energy band. In Figure \ref{fig:Cases}, we apply our dense CSM shell SBO model to fit the X-ray lightcurve of the blackbody component and the observed temperature evolution of XRF\,060218. The model provides a good fit to both the lightcurve and temperature evolution, with posterior constraints of {$r_{\rm out}=1400^{+250}_{-130}\,R_\odot$, $\chi=0.0784^{+0.0085}_{-0.0098}$, $M_{\rm CSM}=2.12^{+0.94}_{-0.40}\times10^{-3}\,M_\odot$, $v_{\rm sh}=0.1435^{+0.003}_{-0.003}\,c$, $t_{\rm first}=450^{+220}_{-50}\,{\rm s}$, $f_L=0.32^{+0.15}_{-0.09}$, and $f_T=0.128^{+0.059}_{-0.052}$, respectively. The inferred optical depths for the entire CSM shell and mass at a SBO radius are $\tau_{\rm CSM}=7.55^{+0.88}_{-0.40}$ and $\tau_{\rm SBO}=6.97^{+0.15}_{-0.15}$, respectively, while the inner and SBO radii are $r_{\rm in}=1290^{+230}_{-130}\,R_\odot$ and $r_{\rm SBO}=1300^{+230}_{-130}\,R_\odot$, respectively. } We note that we do not model the gamma-ray prompt emission and subsequent X-ray afterglow emission, which may originate from a physically distinct component from the thermal emission, such as a usual gamma-ray burst jet \citep{Campana2006,Waxman2007} or repeated bulk Compton scattering of SBO thermal photons by the shock \citep{Wang2007}.

\subsection{XRO\,080109}

XRO\,080109 was serendipitously discovered by {\em Swift}/XRT during observations of the galaxy NGC\,2770, at a luminosity distance of $D_{\rm L}=27\,{\rm Mpc}$. Unlike the other two events, XRO\,080109 was associated with a Type Ibc SN, SN\,2008D \citep{Soderberg2008,Mazzali2008}. The observed X-ray lightcurve in $0.3-10\,{\rm keV}$ peaks at $T_0+60\,{\rm s}$, with a luminosity of a few $10^{43}\,{\rm erg}\,{\rm s}^{-1}$. Here, we extract time-resolved spectra by dividing the XRT data into time intervals, each containing at least 60 accumulated photon counts. Each spectrum is independently fitted with both an absorbed blackbody model and an absorbed power-law model, with the $N_{\rm H}$ fixed to the value obtained from the time-integrated spectral fit. For the majority of the time-resolved spectra, the two models {provide} statistically comparable fits. We adopt the results of the blackbody fits to derive the temperature evolution and the X-ray luminosity.

We fit both the X-ray lightcurve and temperature evolution, and show our fitting result for XRO\,080109 in Figure \ref{fig:Cases}. The posterior constraints are {$r_{\rm out}=37.7^{+0.6}_{-0.6}\,R_\odot$, $\chi=0.850^{+0.010}_{-0.008}$, $M_{\rm CSM}=7.3^{+0.2}_{-0.1}\times10^{-7}\,M_\odot$, $v_{\rm sh}=0.119^{+0.002}_{-0.002}\,c$, $t_{\rm first}=376.0^{+1.3}_{-0.8}\,{\rm s}$, $f_L = 0.003^{+0.007}_{-0.002}$, and $f_T = 0.001^{+0.003}_{-0.001}$, respectively. The inferred optical depths of the entire CSM shell and at shock breakout are $\tau_{\rm CSM}=8.55^{+0.17}_{-0.16}$ and $\tau_{\rm SBO}=8.44^{+0.17}_{-0.13}$, respectively. The corresponding inner and SBO radii are $r_{\rm in}=5.6^{+0.4}_{-0.4}\,R_\odot$ and $r_{\rm SBO}=6.0^{+0.4}_{-0.3}\,R_\odot$. The relatively thick CSM shell inferred for XRO\,080109 ($\chi\sim0.85$) may challenge the constant shock velocity assumption, as the shock could decelerate significantly depending on the density profile and dynamics of the impacting SN ejecta. This effect is not included in our model and may introduce additional uncertainties into the inferred parameters. } Although the model can briefly reproduce the observed X-ray lightcurve and temperature evolution, the inferred CSM parameters suggest that the surrounding material may not originate from a geometrically thin shell. Converting the averaged CSM density into an equivalent mass-loss rate gives $\dot{M}\sim10^{-3}\,M_\odot\,{\rm yr}^{-1}$ by assuming a classical wind velocity of $v_{\rm w}=1000\,{\rm km\,s^{-1}}$, which is much higher than the typical wind mass-loss rates of Wolf-Rayet stars \citep[e.g.,][]{Nugis2000}. This may suggest an enhanced mass loss shortly before the explosion or binary interactions. 

\section{Summary and Discussions}

In this {\em Letter}, we systematically investigate X-ray SBO and post-breakout cooling emission produced when SN ejecta collide with and break out from a dense CSM shell ejected before the explosion. We develop a semi-analytic model to describe the lightcurve and temperature evolution of the X-ray emission, and explore the dependence of the observables on the model parameters.

We find that the proposed model can reproduce the observed properties of the currently known soft X-ray SBO events. Furthermore, our modeling reveals a diversity in the properties of the CSM shells among these events, particularly in their outer radii, shell widths, and CSM masses. These dense CSM shells may originate from enhanced mass-loss episodes of the stripped-envelope progenitor systems shortly before the explosion, such as wave-driven outbursts \citep[e.g,][]{Quataert2012,Fuller2018,Leung2021,Wu2021,Wu2022b} or binary interactions \citep[e.g.,][]{Tauris2015,Wu2022}. Future discoveries of more soft X-ray SBO events will provide a unique opportunity to probe the poorly understood pre-SN mass-loss history of massive stars.

\begin{acknowledgments}
{We thank an anonymous referee for constructive suggestions, Guang-Lei Wu for valuable discussions and comments.} JPZ, YYH, and BZ acknowledge the Research Talent Hub for ITF project (RTH-ITF) (Project No. PiH/270/25GS) from the Innovation Technology Commission of Hong Kong SAR. YYW acknowledges the National Natural Science Foundation of China (grant No. 12393811). ZL acknowledges the National Natural Science Foundation of China (No. 12575116).
\end{acknowledgments}

\begin{contribution}
JPZ, YHY, and BZ initiated the project. All authors  contributed to the development of the model and the writing of the manuscript.
\end{contribution}

\software{\texttt{emcee} \citep{ForemanMackey2013}; \texttt{bayspec} \citep{Yin2025}; \texttt{HEAsoft} \citep{HEAsoft2014}; \texttt{procreate} (\url{https://procreate.com})}

\appendix
\twocolumngrid

\section{Comparison with Existing Models}

\begin{figure}[t!]
\centering
\includegraphics[width=1\linewidth, trim = 72 110 105 60, clip]{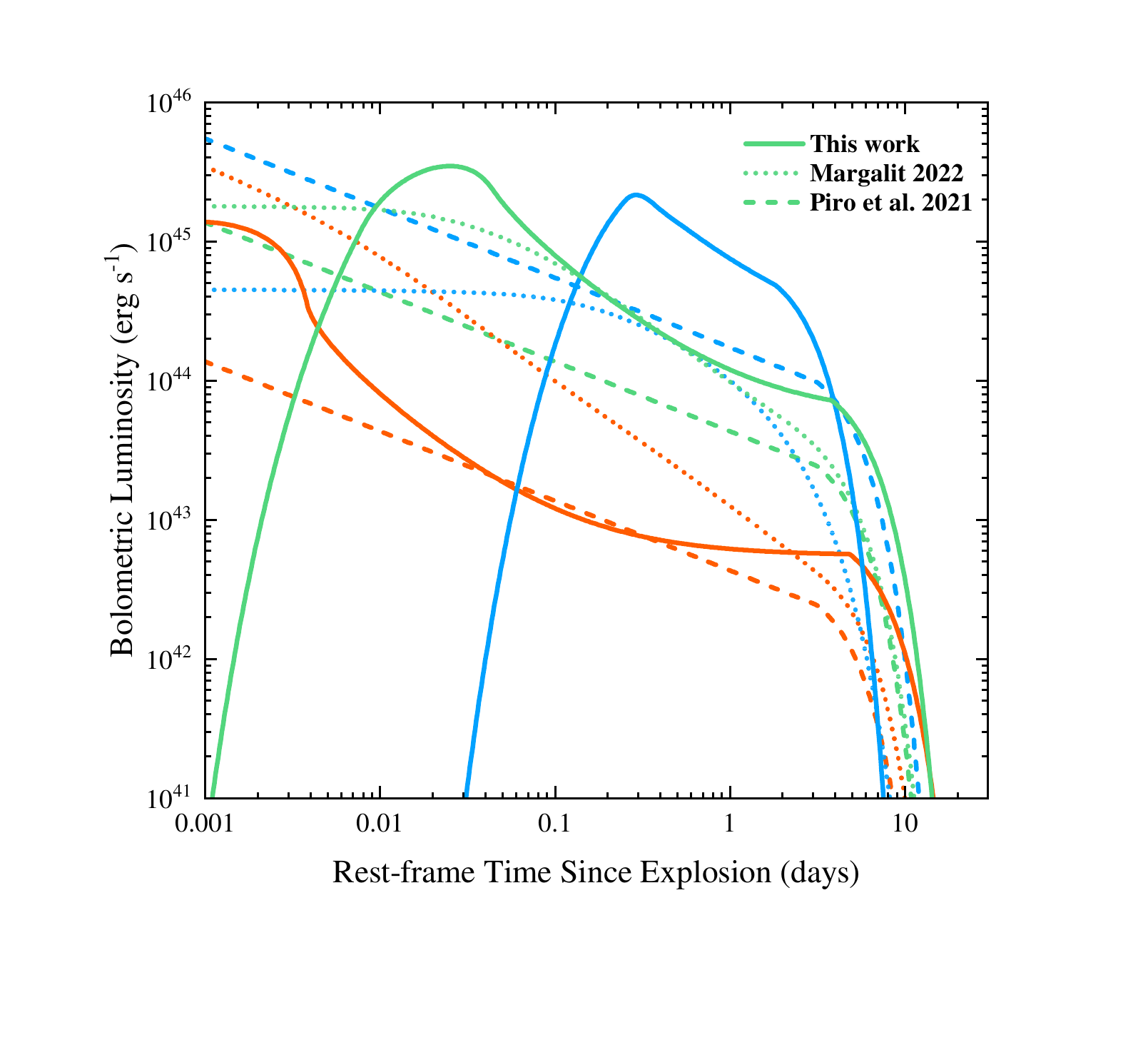}
\caption{{Comparison of our model bolometric lightcurves with those of existing models. The solid lines show our model results, while the dashed and dotted lines correspond to the models of \cite{Piro2021} and \cite{Margalit2022}, respectively. The green lines represent the case with $r_{\rm out}=10^{14}\,{\rm cm}$, $\chi=1/7$, $M_{\rm CSM}=10^{-1}\,M_\odot$, and $v_{\rm sh}=0.033\,c$, while the orange and blue lines correspond to $r_{\rm out}=3.2\times10^{14}\,{\rm cm}$ and $10^{13}\,{\rm cm}$, respectively, with all other parameters unchanged. For consistency, we  note that we adopt an opacity of $\kappa=0.34\,{\rm cm}^2\,{\rm g}^{-1}$ in this comparison.    }}
\label{fig:Comparsion}
\end{figure}

{In Figure \ref{fig:Comparsion}, we compare our model bolometric lightcurves with those predicted by \cite{Piro2021} and \cite{Margalit2022} using three sets of identical parameters. The fiducial model adopts $r_{\rm out}=10^{14}\,{\rm cm}$, $\chi=1/7$, $M_{\rm CSM}=10^{-1}\,M_\odot$, and $v_{\rm sh}=0.033\,c$, while the other two models differ only in $r_{\rm out}$, taking values of $3.2\times10^{14}\,{\rm cm}$ and $10^{13}\,{\rm cm}$. We find that the pre-peak luminosities are broadly comparable among the three models, except that the model with the largest $r_{\rm out}$ is slightly brighter than the other two. Similarly, the lightcurves transition to Gaussian decay at approximately the same time as the diffusion front reaches the inner boundary of the CSM shell. However, our comparison highlights that light-travel effects can significantly influence the lightcurve peak and should not be neglected.}

\section{Markov Chain Monte Carlo Fitting of SBO Events}
\label{app_sec:MCMC}

{Our model includes seven free parameters, for which we adopt the following priors: $\log_{10}(r_{\rm out}/R_\odot)\sim\mathcal{U}(1,4)$, $v_{\rm sh}/c\sim\mathcal{U}(0.01, 0.5)$, $\chi\sim\mathcal{U}(0,1)$, $\log_{10}(M_{\rm CSM}/M_\odot)\sim\mathcal{U}(-8 , 0)$, $t_{\rm first}/{\rm s}\sim\mathcal{U}(0,10^4)$, $f_L\sim\mathcal{U}(0,1)$, and $f_T\sim\mathcal{U}(0,1)$, respectively, where $\mathcal{U}(a,b)$ represents a uniform distribution between $a$ and $b$. Assuming independent Gaussian measurement uncertainties, our log-likelihood function can be expressed as}
\begin{equation}
\begin{split}
\label{equ:Likelihood}
    \ln\mathcal{L}=&-\frac{1}{2}\sum_{i=1}^{N_{L}}\left[\frac{(L_{{\rm X},i}^{\rm obs} - L_{{\rm X},i}^{\rm mod})^2}{s_{L,i}^2}+\ln(2\pi s_{L,i}^2)\right]\\
    &-\frac{1}{2}\sum_{j=1}^{N_{T}}\left[\frac{(T_{{\rm col},j}^{\rm obs} - T_{{\rm col},j}^{\rm mod})^2}{s_{T,j}^2}+\ln(2\pi s_{T,{j}}^2)\right],
\end{split}
\end{equation}
{where $L_{{\rm X},i}^{\rm obs}$ ($T_{{\rm col},j}^{\rm obs}$) and $L_{{\rm X},i}^{\rm mod}$ ($T_{{\rm col},j}^{\rm mod}$) denote the observed and model-predicted X-ray luminosities (color temperatures), respectively, and $N_L$ ($N_T$) is the number of luminosity (temperature) data. The effective uncertainties are defined as 
$s_{L,i}=[\sigma_{L,i}^2+(f_LL_{{\rm X},i}^{\rm obs})^2]^{1/2}$ and
$s_{T,j}=[\sigma_{T,j}^2+(f_TT_{{\rm col},j}^{\rm obs})^2]^{1/2}$,
where $\sigma_{L,i}$ and $\sigma_{T,j}$ are the corresponding
measurement uncertainties.}

\begin{figure*}[t!]
\centering
\includegraphics[width=1\textwidth]{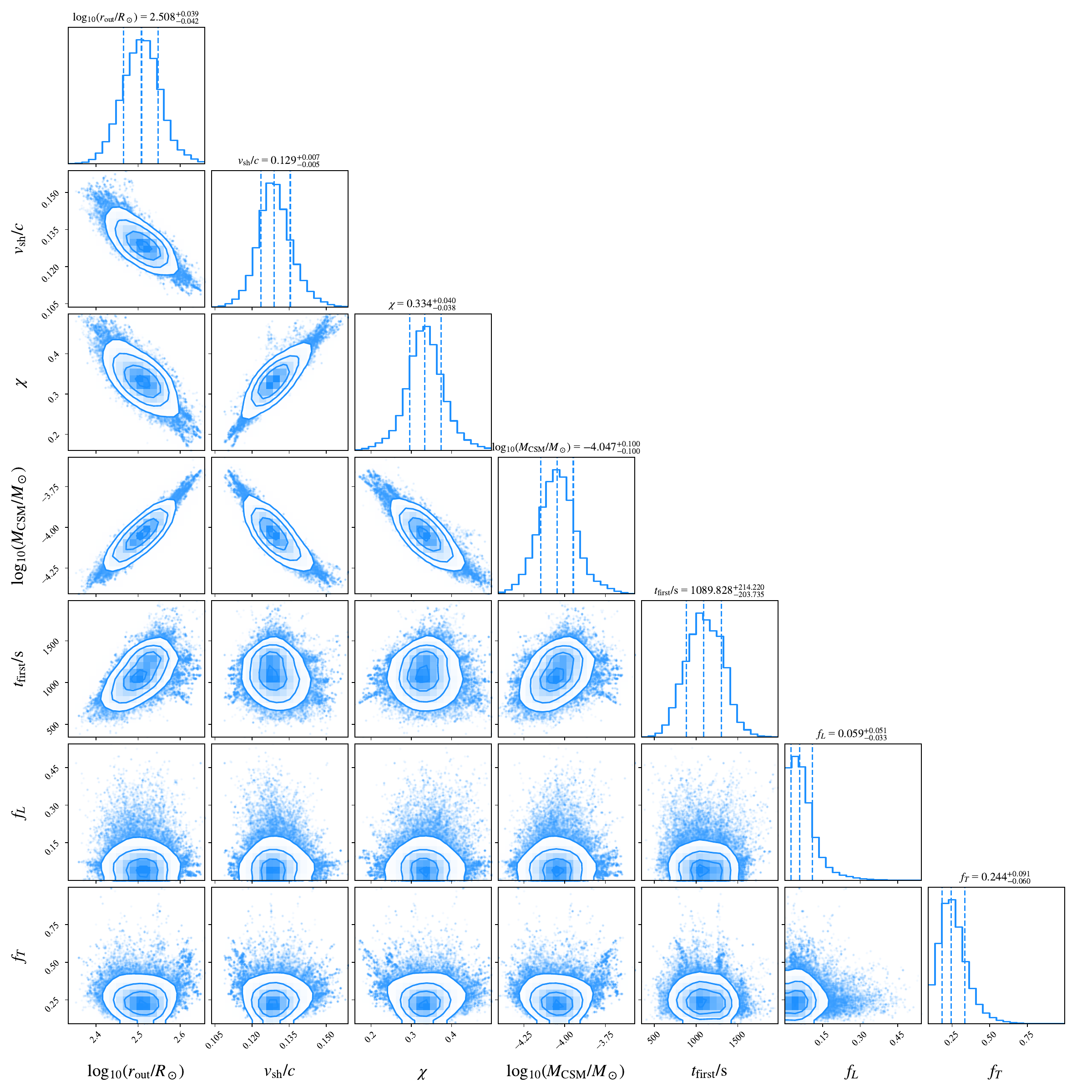}
\caption{Example posteriors for our SBO model fit to EP260321a. Medians and $1\sigma$ credible regions are labeled.  }
\label{fig:Corner}
\end{figure*}

{In Figure \ref{fig:Corner}, we show the posterior
distributions of the parameters for EP260321a as an example, with the corresponding lightcurve and temperature fits presented in Figure \ref{fig:Cases}, to illustrate the convergence of our lightcurve fitting. The posteriors are well behaved and unimodal, indicating good convergence of the Markov Chain Monte Carlo sampling. The inferred posterior constraints for EP260321a are  $r_{\rm out}=322^{+30}_{-30}\,R_\odot$, $v_{\rm sh}=0.129^{+0.007}_{-0.005}\,c$, $\chi=0.334^{+0.040}_{-0.038}$, $M_{\rm CSM}=9.0^{+2.8}_{-1.3}\times10^{-5}\,M_\odot$, $t_{\rm first}=1089^{+214}_{-203}\,{\rm s}$, $f_L = 0.058^{+0.051}_{-0.033}$, and $f_T = 0.244^{+0.091}_{-0.060}$, respectively. We assess goodness of fit using posterior predictive of our likelihood of Equation (\ref{equ:Likelihood}), including $f_L$ and $f_T$. Defining $Q$ as the sum of squared residuals weighted by the total variances, we obtain $Q=119.9^{+8.7}_{-11.1}$ and $Q/(n_{\rm data}-n_{\rm para})=1.82^{+0.13}_{-0.17}$, where $n_{\rm data}=73$ and $n_{\rm para}=7$ denote the numbers of data points and fitted parameters, respectively. The joint residual-sum-of-squares check yields a posterior predictive $p$-value of 0.027. Thus, we suggest that the model captures the broad luminosity and temperature evolution.   }

\bibliography{ms}{}
\bibliographystyle{aasjournalv7}

\end{document}